\documentclass[12pt]{article}
\usepackage{epsfig}
\usepackage{comment}
\usepackage{latexsym}
\usepackage{color}
\usepackage{amsmath}
\newcommand{\mysquare}[0]{\raise-.2ex\hbox{{\Large$\Box$}}}
\def\lsim{\mathrel{\rlap {\raise.5ex\hbox{$ < $}}
{\lower.5ex\hbox{$\sim$}}}}
\def\gsim{\mathrel{\rlap {\raise.5ex\hbox{$ > $}}
{\lower.5ex\hbox{$\sim$}}}} \topmargin -1.5cm \textheight=22.5cm \textwidth=16.5cm
\catcode`\@=11
\newcount\hour
\newcount\minute
\newtoks\amorpm
\hour=\time\divide\hour by60 \minute=\time{\multiply\hour by60 \global\advance\minute by-\hour}
\edef\standardtime{{\ifnum\hour<12 \global\amorpm={am}%
        \else\global\amorpm={pm}\advance\hour by-12 \fi
        \ifnum\hour=0 \hour=12 \fi
        \number\hour:\ifnum\minute<10 0\fi\number\minute\the\amorpm}}
\edef\militarytime{\number\hour:\ifnum\minute<10 0\fi\number\minute}
\def\draftlabel#1{{\@bsphack\if@filesw {\let\thepage\relax
   \xdef\@gtempa{\write\@auxout{\string
      \newlabel{#1}{{\@currentlabel}{\thepage}}}}}\@gtempa
   \if@nobreak \ifvmode\nobreak\fi\fi\fi\@esphack}
        \gdef\@eqnlabel{#1}}
\def\@eqnlabel{}
\def\@vacuum{}
\def\draftmarginnote#1{\marginpar{\raggedright\scriptsize\tt#1}}
\def\draft{\oddsidemargin -.2truein
        \def\@oddfoot{\sl preliminary draft \hfil
        \rm\thepage\hfil\sl\today\quad\militarytime}
        \let\@evenfoot\@oddfoot \overfullrule 3pt
        \let\label=\draftlabel
        \let\marginnote=\draftmarginnote
   \def\@eqnnum{(\theequation)\rlap{\k

 ern\marginparsep\tt\@eqnlabel}%
\global\let\@eqnlabel\@vacuum}  }

\newcommand{\be}[0]{\begin{equation}}
\newcommand{\ee}[0]{\end{equation}}
\newcommand{\ba}[0]{\begin{eqnarray}}
\newcommand{\ea}[0]{\end{eqnarray}}

\def\bs{\begin{subequations}}
\def\es{\end{subequations}}
\def\d{\partial}

\def\thebibliography#1{%
\vskip 0.5cm \centerline{\bf \Large References}
\list{%
[\arabic{enumi}]}{\settowidth\labelwidth{[#1]} \leftmargin\labelwidth \advance\leftmargin\labelsep
\usecounter{enumi}}
\def\newblock{\hskip .11em plus .33em minus .07em}
\sloppy\clubpenalty4000\widowpenalty4000 \sfcode`\.=1000\relax}

\renewcommand{\theequation}{\arabic{section}.\arabic{equation}}

\renewcommand{\section}{\setcounter{equation}{0}\@startsection
{section}{1}{0mm}{-\baselineskip}{0.5\baselineskip} {\normalfont\Large\bfseries}}

\renewcommand{\subsection}{\@startsection
{subsection}{2}{0mm}{-\baselineskip}{0.5\baselineskip} {\normalfont\large\bfseries}}

\renewcommand{\subsubsection}{\@startsection
{subsubsection}{3}{0mm}{-\baselineskip}{0.5\baselineskip} {\normalfont\normalsize\slshape}}

\usepackage{amssymb,amsfonts}
\usepackage{graphicx}
\usepackage{cite}

\def\bc{\begin{center}}
\def\ec{\end{center}}
\def\bea{\begin{eqnarray}}
\def\eea{\end{eqnarray}}

\def\d{\delta}
\def\t{\tau}

\def\k{\kappa}

\def\ad{\dot a}

\def\tt{\tilde t}
\def\hk{\hat k}
\def\kh{\hat k}

\def\where{\quad\mbox{where}\quad}
\def\and{\quad\mbox{and}\quad}

\newcommand{\ie}{{\em i.e. }}

\newcommand{\Z}{\mathbb{Z}}

\newcommand{\tf}{{\tilde f}}
\newcommand{\tg}{{\tilde g}}
\newcommand{\tF}{{\tilde F}}
\newcommand{\tG}{{\tilde G}}

\begin{document}
\begin{titlepage}
\begin{flushright}
LPTENS--07/50, 
CPHT--RR085.0707,
October 2007
\end{flushright}


\begin{centering}
{\bf\huge Thermal/quantum effects and}\\
\vspace{2mm}
{\bf\huge induced superstring cosmologies$^\ast$}\\

\vspace{4mm}
 {\Large Tristan Catelin-Jullien$^{1}$, Costas~Kounnas$^{1}$\\
 Herv\'e~Partouche$^2$ and Nicolaos~Toumbas$^3$}

\vspace{2mm}

$^1$ Laboratoire de Physique Th\'eorique,
Ecole Normale Sup\'erieure,$^\dagger$ \\
24 rue Lhomond, F--75231 Paris cedex 05, France\\
{\em  catelin@lpt.ens.fr, Costas.Kounnas@lpt.ens.fr}

\vskip .1cm

$^2$ {Centre de Physique Th\'eorique, Ecole Polytechnique,$^\diamond$
\\
F--91128 Palaiseau, France\\
{\em Herve.Partouche@cpht.polytechnique.fr}} \vskip .1cm

$^3$ {Department of Physics, University of Cyprus,\\ Nicosia 1678, Cyprus }\\ {\em nick@ucy.ac.cy}
 \vspace{3mm}

{\bf\Large Abstract}

\end{centering}


\noindent We consider classical superstring theories on flat four dimensional space-times,
and where $N=4$ or $N=2$ supersymmetry is spontaneously broken.
We obtain the thermal and quantum corrections at the string one-loop level
and show that the back-reaction on the space-time metric induces a cosmological evolution.
We concentrate on heterotic string models obtained by compactification on a $T^6$ torus and on $T^6/\Z_2$ orbifolds.
The temperature $T$ and the supersymmetry breaking scale $M$ are generated via
the Scherk-Schwarz mechanism on the Euclidean time cycle and on an internal spatial cycle respectively.
The effective field theory corresponds to a no-scale supergravity, where
the corresponding no-scale modulus controls the Susy-breaking scale.
The classical flatness of this modulus is lifted by an effective thermal potential, given by the free energy.
The gravitational field equations admit solutions where $M$, $T$ and the inverse scale factor $1/a$ of
the universe remain proportional. In particular the ratio $M/T$ is fixed during the time evolution.
The induced cosmology is governed
by a Friedmann-Hubble equation involving
an effective radiation term $\sim 1/a^4$ and an effective curvature term $\sim 1/a^2$, whose
coefficients are functions of the complex structure ratio $M/T$.


\vspace{5pt} \vfill \hrule width 6.7cm \vskip.1mm{\small \small \small \noindent $^\ast$\ Research
partially supported by the EU (under the contracts MRTN-CT-2004-005104, MRTN-CT-2004-512194,
MRTN-CT-2004-503369, MEXT-CT-2003-509661), INTAS grant 03-51-6346,
 CNRS PICS 2530, 3059 and 3747,  ANR (CNRS-USAR) contract
 05-BLAN-0079-01 and  INTERREG IIIA Crete/Cyprus.\\
$^\dagger$\ Unit{\'e} mixte  du CNRS et de l'Ecole Normale Sup{\'e}rieure associ\'ee \`a
l'Universit\'e Pierre et Marie Curie (Paris
6), UMR 8549.\\
 $^\diamond$\ Unit{\'e} mixte du CNRS et de l'Ecole Polytechnique,
UMR 7644.}

\end{titlepage}
\newpage
\setcounter{footnote}{0}
\renewcommand{\thefootnote}{\arabic{footnote}}
 \setlength{\baselineskip}{.7cm} \setlength{\parskip}{.2cm}


\setcounter{section}{0}
\section{Introduction}

\noindent It is important to develop a string theoretic framework for studying cosmology. The ultimate goal
of this task is to determine whether string theory can describe basic features of our Universe.
Despite considerable effort towards this direction over the last few years
(see for example \cite{ABS} -- \cite{Skenderis:2007sm}),
still very little is
known about the dynamics of string theory in time-dependent, cosmological settings. The purpose of
this work is to provide a new class of non-trivial string theory cosmological solutions,
where some of the difficult issues can be explored and analyzed concretely.

\noindent At the classical string level, it seems difficult to obtain exact cosmological solutions \cite{Maldacena:2000mw}.
Indeed,
after extensive studies in the framework of superstring compactifications (with or without fluxes),
the obtained results appear to be unsuitable for cosmology. In most cases, the classical ground
states correspond to static Anti-de Sitter or flat backgrounds but not to cosmological ones.
The same situation appears to be true in the effective supergravity theories. Naively, the results obtained in this direction lead to the conclusion that cosmological ground
states are unlikely to be found in superstring theory.

\noindent From our viewpoint this conclusion cannot be correct for two reasons:\\
\indent $\bullet$ The first follows from the fact that already
exact (to all orders in $\alpha'$) cosmological solutions exist, which are
described by a two
dimensional worldsheet conformal field theory based on a gauged Wess-Zumino-Witten
model at negative level $-|k|$:
${SL(2,R)_{-|k|}\over U(1)}\times M$, \cite{kounnas-lust, Nappi, Elitzur}.\\
\indent $\bullet$ The second is that quantum and thermal corrections are
neglected in the classical string/supergravity regime.

\noindent The first class of stringy cosmological models was studied recently in \cite{KTT}, where it was
shown how to define a normalizable wave-function for this class of backgrounds, realizing the Hartle-Hawking
no-boundary proposal \cite{WFU} in string theory. Explicit calculable examples were given for small values of
the level $|k|$. As it was shown in \cite{KTT}, these models are intrinsically thermal with a temperature
below but still close to the Hagedorn temperature.
The disadvantage of small level $|k|$, however, is the absence of a semi-classical
limit with $|k|$ arbitrarily large, which prevents us from obtaining a clean geometrical
picture and studying issues such as back-reaction and particle production in a straightforward way.

\noindent Another direction consists of studying  the ``quantum and thermal cosmological
solutions," which are generated dynamically at the quantum level of string theory \cite{KP,KPThermal}.
Although this study looks to be hopeless and out of any systematic control, it turns out that in
certain cases the quantum and thermal corrections are under control
thanks to the special structure of the underlying effective supergravity theory in its spontaneously broken
supersymmetric phase. An effective field theory study has already been initiated in \cite{KP,KPThermal}.
(See also \cite{Ohta:2004wk, BouhmadiLopez:2006pf}.)

\noindent In order to see how cosmological solutions arise naturally in this context, consider the case of a
supersymmetric flat string background. At finite temperature the thermal fluctuations produce a
non-zero energy density that is calculable perturbatively at the full string level.
The back-reaction on the space-time metric and on certain of the moduli fields
gives rise to a specific cosmological evolution. For temperatures below the Hagedorn temperature,
the evolution of the universe is known to be radiation dominated. (See for instance \cite{Matsuo:1986es,vafabrand}
for some earlier work in this case and \cite{vafabrand} for
a review on string gas cosmology.)

\noindent More interesting cases are those where space-time supersymmetry is spontaneously broken at the string
level either by geometrical \cite{GeometricalFluxes} or non-geometrical fluxes . In the case where the geometrical fluxes are
generated via freely acting orbifolds \cite{Scherk:1978ta} --\cite{akd}, the stringy quantum corrections
are under control in a very
similar way as the thermal ones. The back-reaction of the quantum and thermal corrections on the
space-time metric and the moduli fields results in deferent kinds of cosmologies depending on the
initial amount of supersymmetry $(N=4,N=2,N=1)$.

\noindent In this work we restrict attention to four-dimensional backgrounds with initial $N=4$
or $N=2$ space-time supersymmetry, obtained by toroidal compactification of the heterotic
superstring on $T^6$ and $T^6/\Z_2$-orbifolds.
The spontaneous breaking of supersymmetry is
implemented via freely acting orbifolds (as in \cite{Scherk:1978ta} --\cite{akd}).
The quantum and thermal corrections are determined
simultaneously by considering the Euclidean version of the model where all coordinates are
compactified: $S_T^1\times T^3$ (for the four-dimensional space-time part) $\times M^6$ (for the
internal manifold).
Apart from being interesting in their own right, these examples may give us useful hinds on how to handle
the phenomenologically more relevant $N=1$ cases.
The $N=1$ cases will be studied elsewhere.

\noindent The thermal corrections are implemented by introducing a coupling of the
space-time fermion number
$Q_F$ to the string momentum and winding numbers associated to the Euclidean time cycle $S_T^1$.
The breaking of supersymmetry is generated by a similar coupling of an internal $R$-symmetry charge
$Q_R$ to the momentum and winding numbers associated to an internal spatial cycle
$S_M^1$, e.g. the $X_5$ coordinate cycle.

\noindent We stress here, that the thermal and supersymmetry breaking couplings correspond to string theoretic
generalizations of Scherk-Schwarz compactifications.
Two very special mass scales appear both associated
with the breaking of supersymmetry. These are the temperature scale $T\sim 1/(2\pi R_0)$ and the
supersymmetry breaking scale $M\sim 1/(2\pi R_5)$, with $R_0$ and $R_5$ the radii of the Euclidean time
cycle, $S_T^1$, and of the internal spatial cycle, $S_M^1$, respectively. The initially degenerate mass levels
of bosons and fermions split by an amount proportional to $T$ or $M$, according
to the charges $Q_F$ and $Q_R$. This mass splitting is the signal of supersymmetry
breaking and gives rise to a non-trivial free energy density, which incorporates simultaneously the thermal
corrections and quantum corrections due to the supersymmetry breaking boundary conditions along the spatial
cycle $S_M^1$.

\noindent At weak coupling, the free energy density
can be obtained from the one-loop Euclidean string partition function \cite{Kounnas:1988ye} --\cite{KounnasRostand}.
The perturbative string
amplitudes are free of the usual ultraviolet ambiguities that plague a field theoretic approach towards
quantum gravity and cosmology. For large enough $R_0$, $R_5$, the Euclidean system is also free of tachyons --
the presence of tachyons would correspond to infrared instabilities,
driving the system towards a phase transition \cite{Atick:1988si} --\cite{akd}.
Therefore, the corresponding energy density and pressure can be determined unambiguously, and we can
use them as sources in Einstein's equations to obtain non-trivial cosmological solutions. This perturbative approach
breaks down near the initial space-like singularity. We speculate whether this breakdown of perturbation theory can be
associated with an early universe phase transition.

\noindent The paper is organized as follows.
Section 2 is mainly a review, where we also fix most of our notations and conventions.
We first consider the four-dimensional
heterotic string models at finite temperature.
We obtain the one-loop thermal partition function at the full string
level, and then we discuss the effective field theory limit at large radius $R_0$.
We also review the analogous computation
of the one loop string partition function at zero temperature and in the case where Susy-breaking boundary
conditions are placed along the internal spatial cycle $S^1_M$,  \cite{Scherk:1978ta} --\cite{akd}.
In the large radius limit, the Einstein frame
effective potential is proportional to the fourth power of the gravitino mass scale,
and it can be positive or negative depending on the choice of the Susy-breaking
operator $Q_R$.

\noindent In section 3, we consider the case where thermal and quantum corrections
due to the supersymmetry breaking
are present simultaneously. For the simplest choice $Q_R=Q_F$,
the corresponding one-loop string partition function is invariant under the
$T \leftrightarrow M$ exchange, manifesting the underlying temperature/gravitino mass scale
duality of the models.
This duality is broken by the other allowable choices for the Susy-breaking
operator $Q_R$, which we classify for both the
$N=4$ and the $N=2$ orbifold cases.

\noindent In the large radii $R_0, R_5$ limit, the pressure consists of two pieces:
the purely thermal part which scales as
$n^*_TT^4$, with the coefficient
$n^*_T$ being the number of all massless boson/fermion pairs in the initially supersymmetric theory, and another
potential-like piece which scales as $n^*_VM^4$ and with the coefficient $n^*_V$ being positive or negative
depending on the choice of the operator $Q_R$.
In both pieces, the rest of the dependence on the scales $T$ and $M$ can be expressed neatly in terms of non-holomorphic
Eisenstein series of order $5/2$ whose variable is the complex structure-like ratio $M/T$.
In addition, we incorporate the effects of small, continuous Wilson line deformations in our computation.
Wilson
lines along any of the internal spatial cycles, other than $S^1_M$, introduce new mass scales, and
pieces proportional to $\sim T^2$ and $\sim M^2$ arise in the effective thermodynamic quantities.

\noindent In section 4 we present our ansatz for the induced cosmological solutions. These are homogeneous
and isotropic cosmologies for which the Susy-breaking scales $T$ and $M$ as well as the inverse of
the scale factor $1/a$ evolve the same way in time, and so the ratio of any two of these quantities
is constant.
The form of this ansatz is dictated by the scaling properties of the effective energy
density and pressure. The compatibility of the gravitational field equations with the equation of motion
of the scalar modulus controlling the size of the gravitino mass scale fixes the ratio $M/T$. By solving
the compatibility equations numerically, we find that
in the absence of Wilson lines along $S^1_M$,
 non-trivial four dimensional solutions exist when $n^*_V$
is negative and the ratio $|n^*_V|/n^*_T$ is small enough. These conditions are satisfied by various models we
describe explicitly in the paper. When we include Wilson lines along $S_M^1$, the value of
the ratio $M/T$ for some of the solutions can be large or small, and so we can have models
with a hierarchy for the scales $M$ and $T$.

\noindent Having solved the compatibility equations, the time-dependence of the system is governed solely by the
familiar Friedmann-Hubble equation. There is a radiation term, $c_r/a^4$,
whose coefficient $c_r$ is positive in our examples. An effective curvature term, $-\hat k /a^2$,
can be generated by turning on Wilson line deformations.
The sign of $\hat k$ can be a priori  positive or negative, depending on the model. When we turn on the kinetic terms of some
of the extra flat moduli, we generate an additional term that scales as $c_m/a^6$ (with $c_m$ positive).

\noindent In section 5, we solve the Friedmann-Hubble equation for the various possible cases,
and we elaborate on the properties of the
cosmological solutions:\\
\indent $\bullet$ When $c_r>0$, we have standard hot big bang cosmologies with an intermediate radiation dominated
era. The late time behavior is governed by the spatial curvature of the models.\\
\indent $\bullet$ We also consider a priori possible exotic models characterized by $c_r<0$.
A big bang occurs when $c_m>0$. The cosmological evolution always ends with a big crunch when $\hk\ge 0$.
The case $\hk < 0$ however is more interesting.
It involves either a first or second order phase transition between the big bang cosmology
and a linearly expanding universe.
The first case corresponds to a tunneling effect involving a gravitational instanton,
while the transition is smooth in the second case.
If the first order transition does not occur, the universe ends in a big crunch. \\
We finish with our conclusions and directions for future research.

\section{Thermal and quantum corrections in heterotic backgrounds }
Our starting point is the class of four dimensional string backgrounds obtained by toroidal
compactification of the heterotic string on $T^6$ and $T^6/\Z_2$ orbifolds. Initially the amount of
space-time supersymmetry is $N_4=4$ for the case of compactification on the $T^6$ torus and $N_4=2$
for the orbifold compactifications, and the four dimensional space-time metric is flat. Space-time
supersymmetry is then spontaneously broken by introducing Scherk-Schwarz boundary conditions on an
internal spatial cycle and/or by thermal corrections. Due to the supersymmetry breaking, the
one-loop string partition function is non-vanishing, giving rise to an effective potential. Our aim
is to determine the back-reaction on the initially flat metric and moduli fields.

\noindent At the one-loop level, the four dimensional string frame effective action is given by \be S=\int
d^4 x\sqrt{-\det g}\left( e^{-2\phi}({1 \over 2}R+2\partial_{\mu}\phi\partial^{\mu}\phi+\cdots)-{\rm
\cal V}_{\rm String}\right),\ee where $\phi$ is the $4d$ dilaton field and the ellipses stand for
the kinetic terms of other moduli fields (to be specified later). At zero temperature, the
effective potential ${\rm \cal V}_{\rm String}$ can be obtained from the one-loop Euclidean string
partition function as follows: \be {Z \over V_4}= -{\rm \cal V}_{\rm String}\, , \ee with $V_4$ the $4d$
Euclidean volume. The absence of a dilaton factor multiplying the potential term in the action is
due to the fact that this arises at the one loop level.

\noindent At finite temperature, the one-loop Euclidean partition function determines the free energy density
and pressure to this order \be {Z \over V_4}= -{ \cal F}_{\rm String}=P_{\rm String}.\ee The
subscript indicates that these densities are defined with respect to the string frame metric. The
relevant Euclidean amplitude incorporates simultaneously the thermal corrections and quantum
corrections which arise from the spontaneous breaking of supersymmetry and which are present even
at zero temperature.

\noindent In order to determine the back-reaction of the (thermal and/or) quantum corrections, it is
convenient to work in the Einstein frame where there is no mixing between the metric and the
dilaton kinetic terms. We define as usual the complex field $S$, \be S=e^{-2\phi}+i\chi, \ee where $\chi$
is the axion field. Then after the Einstein rescaling of the metric, the one loop effective action
becomes: \be S=\int d^4x \sqrt{-\det g}\left[{1\over 2}R -g^{\mu\nu}~K_{I \bar J}
 ~\partial_{\mu}\Phi_I \partial_{\nu} \bar\Phi_{\bar J}~
-{1\over s^2}~{\rm \cal V}_{\rm String} (\Phi_I, \bar\Phi_{\bar I})\right], \ee where  $K_{i \bar
\jmath}$ is the metric on the scalar field manifold $\{\Phi_I\}$, which is parameterized by various
compactification moduli including the field $S$. This manifold includes also the main moduli fields
$T_I,\, U_I,~I=1,2,3$, which are the volume and complex structure moduli of the three internal
$2$-cycles respectively. We notice that in the Einstein frame the effective potential, ${\rm \cal
V}_{\rm String} $, is rescaled by a factor $1/ s^2$, where $s={\rm Re} (S)=e^{-2\phi}$. Taking this
rescaling into account, we have \be{\rm \cal V}_{\rm Ein}={1\over s^2}{\rm \cal V}_{\rm String}.
\label{VEinString}\ee This relation will be crucial for our work later on. (We will always work in
gravitational mass units, with $ M_G ={1\over \sqrt{ 8 \pi G_N}}=2.4\times10^{18}$ GeV).

\noindent Keeping only the main moduli fields $\{S,T_I,U_I\}$, their kinetic terms are determined in terms of
the K\"alher potential $K$ \cite{PhiKineticTermes, noScale}:
\be K=-\log ~(S+\bar S) -\sum_I~\log ~(T_I+\bar T_I)-\sum_I~\log
~(U_I+\bar U_I)\ee with $K_{I \bar J}=\partial_I\partial_{\bar J}K$. The classical superpotential
depends on the way supersymmetry is broken.
Generically string backgrounds with spontaneously broken
supersymmetry are flat at the classical level due to the no-scale structure of the effective
supergravity theory \cite{noScale}. Once the thermal and/or quantum corrections are taken into account, we obtain
in some cases interesting cosmological solutions.

\subsection{Heterotic supersymmetric backgrounds at finite
temperature}

In order to fix our notations and conventions, we first consider the case of an exact
supersymmetric background at finite temperature \cite{Atick:1988si} --\cite{akd}.
For definiteness we choose the heterotic string
with maximal space-time supersymmetry ($N_4=4$). All nine spatial directions as well as the
Euclidean time are compactified on a ten dimensional torus. At zero temperature, the Euclidean
string partition function is zero due to space-time supersymmetry. At finite temperature however
the result is a well defined finite quantity. Indeed, at genus one the string partition function is
given by:
\be Z=\oint_F {d\tau d\bar{\tau}\over 4{\rm Im}\tau~}~{1\over 2}\sum_{a,b}
(-)^{a+b+ab}~\theta\left[^a_b\right]^4~{\Gamma_{(10,26)}\left[^a_b\right] \over \eta(\tau)^{12}~
{\bar\eta}(\bar\tau)^{24}}~, \ee
where $\Gamma_{(10,26)}\left[^a_b\right]$ is a shifted Narain
lattice (which we specify more precisely below). The non-vanishing of the partition function is due
to the non-trivial coupling of the lattice to the spin structures $(a,b)$. Here, the argument
$a$
is zero for space-time bosons and one for space-time fermions.
The spin/statistics connection and modular invariance require that the unshifted $\Gamma_{(1,1)}$
sub-lattice of the Euclidean time cycle \be \Gamma_{(1,1)}\equiv\sum_{m,n}~R_0({\rm
Im}\tau)^{-{1\over2}}~e^{-\pi R_0^2{|m+n\tau|^2\over {\rm Im}\tau}}~\ee be replaced as follows:
 \be
\Gamma_{(1,1)}\longrightarrow \sum_{m,n}R_0({\rm Im}\tau)^{-{1\over2}}~e^{-\pi
R_0^2{|m+n\tau|^2\over {\rm Im} \tau}}~e^{i\pi(ma+nb+mn)}~. \ee

\noindent Redefining \be m\rightarrow 2m +g,
~~~~n\rightarrow 2n+h, \ee where $g, h$ are integers defined modulo $2$, and introducing the
notation $\Gamma_{(1,1)}\left[^h_g\right]$ for a shifted lattice,
 \be
\Gamma_{(1,1)}\left[^h_g\right]= \sum_{m,n}R_0({\rm Im}\tau)^{-{1\over2}}~e^{-\pi
R_0^2{|2m+g+(2n+h)\tau|^2\over {\rm Im} \tau}}~, \ee the thermal partition function takes the form:
\be Z=\oint_F {d\tau d\bar{\tau}\over 4{\rm Im}\tau~}~{1\over 2}\sum_{(a,b),(h,g)} ~(-)^{ga+hb
+hg}~ (-)^{a+b+ab}~\theta\left[^a_b\right]^4~{\Gamma_{(9,25)}\Gamma_{(1,1)}\left[^h_g \right] \over
\eta(\tau)^{12}~ {\bar\eta}(\bar\tau)^{24}}~. \ee Defining ${\hat a} =a-h$ and ${\hat b} = b-g$ and
using the Jacobi identity \be {1\over 2}\sum_{({\hat a},{\hat b})}~(-)^{{\hat a}+{\hat b}+{\hat a}
{\hat b}}~\theta\left[^{\hat a+h}_{\hat b +g}\right]^4 =-\theta\left[^{1+h}_{1+g} \right]^4~,\ee we
obtain \be Z=\oint_F {d\tau d\bar{\tau}\over 4{\rm Im}\tau~}~\sum_{(h,g)}
~-(-)^{g+h}~\theta\left[^{1+h}_{1+g} \right]^4 \Gamma_{(1,1)}\left[^h_g \right] {\Gamma_{(9,25)}
\over \eta(\tau)^{12}~ {\bar\eta}(\bar\tau)^{24}}~. \ee The temperature in string frame is given by
$T_{\rm String}=1/(2\pi R_0)$.

\noindent Since our aim is the study of induced cosmological solutions in $3+1$ dimensions, we consider the
case for which the radii of three spatial directions are very large: $R_x=R_y=R_z\equiv R\gg 1$. In
this case the three dimensional spatial volume factorizes
 \be \Gamma_{(3,3)}\cong R^{3}~({{\rm
Im}\tau})^{-{3\over 2}}= {V_3 \over (2\pi)^3}~({{\rm Im}\tau})^{-{3\over 2}}.
\ee
 Using the expression
for the $\Gamma_{(1,1)}\left[^h_g\right]$ shifted lattice we obtain:
$$Z~=~-(2\pi R_0)V_3~{\rm \cal F}_{\rm String}~=~V_4~P_{\rm String}~
$$
\be =~-{V_4 \over (2\pi)^4}\oint_F {d\tau d\bar{\tau}\over ~{4\rm
Im}\tau^3~}\sum_{(n,m),(h,g)}(-)^{g+h}~e^{-\pi R_0^2{|2m+g+(2n+h)\tau|^2\over {\rm Im}
\tau}}~\theta\left[^{1+h}_{1+g} \right]^4~ {\Gamma_{(6,22)} \over \eta(\tau)^{12}~
{\bar\eta}(\bar\tau)^{24}}~,\ee where $V_4=(2\pi R_0)V_3$ is the four dimensional space-time
volume, ${\rm \cal F}_{\rm String}$ the
free energy density and $P_{\rm String}$ the pressure in string frame.

\noindent Before we proceed further, we make some comments:\\
\indent $\bullet$ The sector $(h,g)=(0,0)$ gives zero contribution. This is due to the fact that we started
with a supersymmetric background.\\
\indent$\bullet$ In the odd winding sector, $h=1$, the partition function
diverges when $R_0$ is between the Hagedorn radius $R_H=(\sqrt{2}+1)/2$ and its dual $1/R_H$:
${1\over R_H} < R_0 < R_H $. The divergence is due to a winding state that is tachyonic when $R_0$
takes values in this range, and it signals a phase transition around
the Hagedorn temperature \cite{Atick:1988si} --\cite{akd}. In
this paper we study the regime $R_0 > R_H$, where there is no tachyon and the odd winding sector is
exponentially suppressed. The high temperature regime and the cosmological consequences of the
phase transition will be examined in future work \cite{malakes2}. \\ 
\indent $\bullet$ When $R_0\gg 1$, the contributions of the
oscillator states are also exponentially suppressed, provided that the moduli parameterizing the
internal $\Gamma_{(6,22)}$ lattice are of order unity.

\subsection{The effective field theory in the large $R_0$ limit}
As we already mentioned, the $h=1$ sector of the theory gives exponentially suppressed
contributions of order ${\rm {\cal O}}(e^{-R_0^2})$. Also, the $(h,g)=(0,0)$ sector vanishes due to
supersymmetry. Thus for large $R_0$, only the sector $(h,g)=(0,1)$ contributes significantly. Using
the identity: \be \Gamma_{(1,1)}(R_0)=\Gamma_{(1,1)}\left[^0_0\right]+\Gamma_{(1,1)}\left[^0_1\right]
+\Gamma_{(1,1)}\left[^1_0\right]+\Gamma_{(1,1)}\left[^1_1\right] \ee and neglecting the $h=1$
sectors, we may replace \be \Gamma_{(1,1)}\left[^0_1\right]\rightarrow
\Gamma_{(1,1)}(R_0)-\Gamma_{(1,1)}\left[^0_0\right]=\Gamma_{(1,1)}(R_0)-{1\over 2}\Gamma_{(1,1)}(2R_0)\ee
in the integral expression for $Z$. For each lattice term we decompose the contribution in modular
orbits: $(m,n)=(0,0)$ and $(m,n)\neq(0,0)$. For $(m,n)\neq(0,0)$, the integration over the
fundamental domain is equivalent with the integration over the whole strip but with $n=0$. The
$(0,0)$ contribution is integrated over the fundamental domain. Now the $(0,0)$ contribution of
$\Gamma_{(1,1)}(R_0)$ cancels the one of ${1\over 2}\Gamma_{(1,1)}(2R_0)$, and we are left with the
integration over the whole strip: \be Z =~{V_4 \over (2\pi)^4}\int_{||} {d\tau d\bar{\tau}\over
~{4\rm Im}\tau^3~}\sum_{m}~e^{-\pi R_0^2{(2m+1)^2\over {\rm Im} \tau}}~\theta\left[^{1}_{0}
\right]^4~ {\Gamma_{(6,22)} \over \eta(\tau)^{12}~ {\bar\eta}(\bar\tau)^{24}}~.\ee

\noindent The integral over $\tau_1$ imposes the left-right level matching condition. The left-moving part
contains the ratio \be {\theta\left[^{1}_{0} \right]^4 \over \eta^{12}} =2^4+{\rm \cal
O}(e^{-\pi\tau_2}), \ee which implies that the lowest contribution is at the massless level. Thus
after the integration over $\tau_1$ $(\tau_2\equiv t)$, the partition function takes the form \be Z
=~{V_4 \over (2\pi)^4} \int_0^{\infty }{dt \over 2t^3~}\sum_{m}~e^{-\pi R_0^2{(2m+1)^2\over t}}~
\left(2^4~ D_0  + \sum D(\mu)~e^{-\pi t\mu^2}\right), \ee where $D(\mu)$ denotes the multiplicity
of the mass level $\mu$ and $2^4~ D_0$ is the multiplicity of the massless level. Changing the
integration variable by setting $t=\pi R_0^2 (2m +1)^2~x$, we have: \be Z =~{V_4\over \pi^2 (2\pi
R_0)^4}~\sum_{m}{1\over (2m+1)^4} \int_0^{\infty }{dx \over 2x^3~}~e^{-{1\over x}}~ \left(2^4~ D_0
+ \sum D(\mu)~e^{-x\pi^2(2m+1)^2{\mu^2 R^2_0}}\right). \ee Now the second term in the parenthesis
is exponentially suppressed when the masses $\mu$ are of order (or close) to the string oscillator
mass scale. This will be the case when all of the internal radii and the Wilson-line moduli of the
$\Gamma_{(6,22)}$ lattice are of order unity. For this specific case, the partition function
simplifies to
 \be
Z =~2^3~ D_0 {V_4\over \pi^2 (2\pi R_0)^4}~\sum_{m}{1\over (2m+1)^4} ={2^3~ D_0~\pi^2 \over
48}~{V_4\over (2\pi R_0)^4}~= {1\over 3}~{n^* \pi^2 \over 16}~{V_4~T_{\rm String}^4}, \ee where $n^*=2^3 D_0$
is the number of the massless boson/fermion pairs in the theory. The free energy density and
pressure in string frame are given by \be P_{\rm String}=-{\cal F}_{\rm String}={1 \over 3}{n^* \pi^2~ T_{\rm String}^4 \over 16}.
\ee

\noindent In the Einstein frame, energy densities are rescaled by a factor $1/s^2$ as in Eq. 
(\ref{VEinString}). Thus the pressure and free energy density in this frame are given by \be P_{\rm
Ein}=-{\cal F}_{\rm Ein}={1 \over 3}{n^* \pi^2~ T_{\rm String}^4 \over 16~s^2}={1 \over 3}{n^* \pi^2~ T^4
\over 16}, \label{purethermal}\ee where $T=T_{\rm String}/\sqrt{s}$ is the proper temperature in the Einstein
frame. This result is expected from the effective field theory point of view. When only massless
states are thermally excited, the field theory expression for the pressure is given by \be P={1
\over 3}~\left(n_B+{7\over 8}n_F\right){ \pi^2~ T^4 \over 30},\label{purethermalfield}\ee where
$n_B$ and $n_F$ are the numbers of massless bosonic and fermionic degrees of freedom respectively.
When $n_B=n_F=n^*$, as in a supersymmetric theory, we recover Eq. (\ref{purethermal}).

\subsection{Spontaneous breaking of supersymmetry at zero temperature }
In this case we consider the same class of heterotic models, but now the breaking of supersymmetry
arises due to the coupling of the space-time fermion number to the momentum and winding quantum
numbers of an internal spatial cycle \cite{Scherk:1978ta} --\cite{akd}. Since the temperature is taken to be zero, the spin
structures $(a,b)$ do not couple to the quantum numbers of the Euclidean time cycle which will be
taken to be very large. We also consider the case where three additional spatial directions are
large. Following similar steps to the purely thermal case, the partition function is given by \be Z
=~-{V_5 \over (2\pi)^5}\oint_F {d\tau d\bar{\tau}\over ~4{\rm Im}\tau^{7 \over
2}~}\sum_{(n,m),(h,g)}(-)^{g+h}~e^{-\pi R_5^2{|2m+g+(2n+h)\tau|^2\over {\rm Im}
\tau}}~\theta\left[^{1+h}_{1+g} \right]^4~ {\Gamma_{(5,21)} \over \eta(\tau)^{12}~
{\bar\eta}(\bar\tau)^{24}}~,\ee where now $V_5=V_4(2\pi R_5)$ is a five dimensional volume and the
$\Gamma_{(5,21)}$ lattice parameterizes the internal space. Here also, the $h=1$ sectors give
exponentially suppressed contributions ${\rm \cal O}(e^{-R_5^2})$, and the $(h,g)=(0,0)$ sector
vanishes due to supersymmetry. The rest of the steps can be repeated as in the derivation above to
find \be Z =~{V_5 \over (2\pi)^5} \int_0^{\infty }{dt \over 2t^{7 \over 2}~}\sum_{m}~e^{-\pi
R_5^2{(2m+1)^2\over t}}~ \left(2^4~ D_0 + \sum D(\mu)~e^{-\pi t\mu^2}\right), \ee which after the
change of variables $t=\pi R_5^2 (2m +1)^2~x$ gives \be Z =~{V_5 \over \pi^{5\over 2} (2\pi
R_5)^5}~\sum_{m}{1\over |2m+1|^{5}}
 \int_0^{\infty }{dx \over 2x^{7 \over 2}~}~e^{-{1\over x}}~
\left(2^4~ D_0 + \sum D(\mu)~e^{-x\pi^2(2m+1)^2{\mu^2 R^2_5}}\right). \ee For $\mu$ of order unity,
this simplifies to \be Z= 2\left(1- 2^{-{5}}\right)~{\zeta(5)\Gamma\left({5\over
2}\right)\over \pi^{5\over 2}}~ n^*~{V_4\over (2\pi R_5)^4} \label{VN4zeroT} \ee with $n^*
=2^3D_0$.

\noindent This result
was expected from the effective field theory point of view. Indeed in a theory with spontaneously
broken $N=4$ supersymmetry, the one loop effective potential receives a non-zero contribution
proportional to the mass super-trace $S{\rm tr} {\rm {\cal M}}^4$, which in turn is proportional to
the fourth power of the gravitino mass. The super-traces $Str{\rm {\cal M}}^n$ vanish for $n<N=4$.
In the example of supersymmetry breaking we examined above, the masses of the states are shifted
according to their spin. For initially massless states, the mass after supersymmetry breaking
becomes : \be M^2_Q \rightarrow {Q_F^2\over R^2_5}.\ee This shows that the string frame gravitino
mass is of order $M_{\rm String}\sim{1/R_5}$ and thus $Str {\rm {\cal M}}^4\sim {c/ R^4_5}$.
Including the contributions from all Kaluza-Klein states, one obtains the result given in formula
(\ref{VN4zeroT}). We obtain for the string frame effective potential:
\be
{\rm \cal V}_{\rm String}
=-{Z \over V_4}=-2\left(1- 2^{-{5}}\right)~{\zeta(5)\Gamma\left({5\over 2}\right)\over
\pi^{5\over 2}}~ n^*~{1\over (2\pi R_5)^4}.
\ee

\noindent In the Einstein frame, we have ${\rm \cal V}_{\rm Ein}={1\over s^2}{\rm \cal V}_{\rm String}$ -- see
Eq. (\ref{VEinString}) -- so that
 \be
  {\rm \cal V}_{\rm Ein}=-2\left(1- 2^{-{5}}\right)~
{\zeta(5)\Gamma({5\over 2})\over \pi^{5\over 2}}~ n^*~{1\over s^2 (2\pi R_5)^4}=-C_V
{1\over (s~t_1u_1)^2} =-C_V~M^4,
\ee
where $t_1={\rm Re} (T_1),~u_1={\rm Re} (U_1)$, and $M=1/(st_1u_1)^{1/2}$ is
the gravitino mass scale in the Einstein frame.

\noindent We stress here that the one loop effective potential depends only on the gravitino mass scale,
which in turn depends only on the product of the $s$, $t_1$ and $u_1$ moduli. This suggests to
freeze all moduli and keep only the diagonal combination
\be
3~\log z= \log s+\log t_1+ \log u_1~.
\ee
 The K\"alher potential of the diagonal modulus $Z$, (with $z={\rm Re} (Z)$), takes the
well known $SU(1,1)$ structure\cite{noScale} \be K=-3\log(Z+\bar Z)\, .
\ee
 This gives rise to the kinetic term
and gravitino mass scale,
\be -g^{\mu\nu}~3{\partial_{\mu} Z
\partial_{\nu} \bar Z \over (Z+\bar Z)^2}, \quad \quad
M^2= 8 e^K ={8\over (Z+\bar Z)^3}\, .
\ee
Freezing ${\rm Im} Z$ and defining the field $\Phi$ by
\be
e^{2\alpha \Phi}= M^2={8\over (Z+\bar Z)^3}\, , \ee one finds the kinetic term \be
-g^{\mu\nu}~3{\partial_{\mu} Z\partial_{\nu} \bar Z \over (Z+\bar Z)^2}= -g^{\mu\nu}~{\alpha^2
\over 3}~\partial_{\mu} \Phi\partial_{\nu} \Phi\, .
\ee
The choice $\alpha^2=3/2$ normalizes
canonically the kinetic term of the modulus $\Phi$. The potential for this particular model is: \be
{\rm \cal V}_{\rm Ein}(\Phi)=-C_V~M^4=-C_V~e^{4\alpha\Phi},~~~~\alpha=\sqrt{3\over2}\, .
\ee

\noindent Observe that in this simple model the sign of the potential is negative. As we now explain, we can
construct models with a positive potential, but with the rest of the dependence on the modulus
$\Phi$ being the same. All we have to do is to couple the momentum and winding numbers of the
Scherk-Schwarz cycle not only to the space-time fermion number but also to another internal charge.
For example consider the $ E_8 \times E_8^{\prime}$ heterotic string on $T^6$ and instead of
coupling just to $Q_F$, we couple to $Q_F+Q_{E_8}+Q_{E_8}^{\prime}$, where $Q_{E_8}$ denotes the
charge of an $E_8$ representation decomposed in terms of $SO(16)$ ones, and similarly for
$Q_{E_8}^{\prime}$. These charges take half integer values for the spinorial representations and
integer values for the others. The initial Susy-breaking co-cycle gets modified as follows \be
(-)^{ag+bh+hg}~\longrightarrow~ (-)^{(a+\bar{\gamma} +{\bar
\gamma}^{\prime})g+(b+{\bar\delta}+{\bar\delta}^{\prime})h+hg}, \ee where as before the argument
$a$ is one for space-time fermions and zero for space-time bosons, and
$(\bar\gamma,{\bar\gamma}^{\prime})=(1,1)$ for the spinorial representations of $SO(16)\times
SO(16)^{\prime}$ and $(0,0)$ for the adjoint representations. This operation breaks explicitly the
$ E_8 \times E_8^{\prime}$ gauge group to $S0(16)\times SO(16)^{\prime}$. Proceeding in similar way
as in the previous example,
one finds:
 \be
 Z= 2\left(1-
2^{-{5}}\right)~{\zeta(5)\Gamma\left({5\over 2}\right)\over \pi^{5\over 2}}~ {\tilde
n}^*~{V_4\over (2\pi R_5)^4},
 \ee
 where
 \be
\label{nv-8}
 {\tilde
 n}^*=2^3\left[~[2]_{X_{2,3}}~+~[6]_{T^6}~+~[120-128]_{E_8}~+~[120-128]_{E_8^{\prime}}~\right]
 =~-2^3\times 8 =~-64.
\ee In the previous example only positive signs appear in the above formula since there is no
coupling of the Scherk-Schwarz lattice quantum numbers to the $ E_8\times E_8^{\prime}$ charges,
giving the value $n^*=2^3\times 504$. The reversing of sign for some representations indicates that
it is for the bosons that the masses are shifted and not for the fermions in the corresponding
multiplet.

\noindent We note that in the $N=4$ case, we cannot change the left-multiplicity since all of the left-moving
R-charges are equivalent as required by symmetry. This however is not true for the $N=2$ and
$N=1$ cases. Consider for instance the class of $N=2$ supersymmetric backgrounds obtained by
compactifying the heterotic string on a ${T^4/ \Z_2}$ orbifold (e.g. the $\Z_2$-orbifold limit of the
$K_3$ CY-compactification). In this class of models
(see for instance \cite{Kiritsis}) four internal supercoordinates are twisted and
the corresponding four internal R-charges are half-shifted. The Euclidean partition function is
given by
$$ \!\!\!\!\!\!\!\!\!\!\!\!\!\!\!\!\!\!Z=\oint_F {d\tau d\bar{\tau}\over 4{\rm Im}\tau~}~ {1\over
4}\sum_{(a,b),(H,G)} (-)^{a+b+ab}~{\theta\left[^a_b\right]^2 \theta\left[^{a+H}_{b+G}\right]^2\over
~\eta(\tau)^{4}}$$ \be\;\;\;\;\;\;\;\;\;\;\;\;\;\;\;\;\;\;\;\;\times~ {\Gamma_{(1,1)}(R_0) ~\Gamma_{(3,3)}({\rm space})\over
\eta(\tau)^{2}~ {\bar\eta}(\bar\tau)^{2}}~ ~Z_{(2,2+n_0)}\left[^0_0\right]
~Z_{(4,4+n_t)}\left[^H_G\right]. \label{N2partition} \ee Here $Z_{(2,2+n_0)}$ is the contribution
of two internal coordinates\footnote{In our notations, the space-time coordinates are $X_{0,\dots, 3}$, while the internal ones are $X_{5,\dots, 10}$.}  ($X_5, X_6$) and $n_0$-right moving world-sheet bosons $\phi_i$. Before
supersymmetry breaking, the corresponding $(2,2+n_0)$-lattice is unshifted. $Z_{(4,4+n_t)}$ stands
for the contribution of four internal coordinates ($X_7, X_8,X_9, X_{10}$) all of which are
$\Z_2$-twisted by $(H,G)$, and $n_t$-right moving world-sheet bosons $\phi_I$ which can be
$\Z_2$-twisted breaking part of the initial gauge group. The $\theta$-function terms come from the
contribution of the left-moving world-sheet fermions. Four of them are $\Z_2$-twisted by $(H,G)$.
The contribution associated to the space-time bosons is when $a=0$, while the one associated to the
space-time fermions is when $a=1$.

\noindent From the above supersymmetric $N=2$ partition function, the thermal partition function is obtained
in a way similar to the $N=4$ example, by the following replacement of the Euclidean time
sub-lattice:
\be \Gamma_{(1,1)}(R_0)~~\longrightarrow~~\Gamma_{(1,1)}\left[^{h_1}_{g_1}\right](R_0)~(-)^{
g_1a+h_1b+h_1g_1}. \ee In the case of Scherk-Schwarz spontaneous supersymmetry breaking, the
partition function can be obtained by a similar replacement of the internal $X_5$ coordinate
lattice, either by utilizing the same operator $Q_F$ \be
\Gamma_{(1,1)}(R_5)~~\longrightarrow~~\Gamma_{(1,1)}\left[^{h_2}_{g_2}\right](R_5)~(-)^{
g_2a+h_2b+h_2g_2} \ee or by utilizing an R-symmetry operator associated to one of the twisted
complex planes \be
\Gamma_{(1,1)}(R_5)~~\longrightarrow~~\Gamma_{(1,1)}\left[^{h_2}_{g_2}\right](R_5)~(-)^{g_2(a+H
)+h_2(b+G)+h_2g_2} \label{Rshifted}. \ee

\noindent These are in fact the only two possibilities involving left-moving R charges since all others are
equivalent choices. However, many other choices exist by utilizing parity-like operators involving
the right moving gauge charges $\sum {\bar\gamma}_i$, as in the explicit example of $SO(16) \times
SO(16)^{\prime}$ spinorial representations we gave above: \be
\Gamma_{(1,1)}(R_5)~~\longrightarrow~~\Gamma_{(1,1)}\left[^{h_2}_{g_2}\right](R_5)~(-)^{g_2(a+H
+\sum {\bar\gamma}_i)+h_2(b+G+\sum {\bar\delta}_i)+h_2g_2}. \label{Qgauge} \ee In the next section
we examine representative examples in the case where thermal and spontaneous Susy breaking operations are present simultaneously.


\section{Thermal and spontaneous breaking of Susy}
The most interesting situation for cosmological applications is
the case where spontaneous supersymmetry breaking and thermal corrections are taken into account
simultaneously.
\subsection{Untwisted sector}
\label{untwisted}

\noindent The untwisted sector of the $N=2$ case, $(H,G)=(0,0)$ in Eq. 
(\ref{N2partition}),\footnote{For $h_1=h_2=0$ (even windings), the sector $(H,G)=(0,1)$ gives zero
net contribution due to the identity $ \frac 12\sum_{a,b} (-)^{a+b+ab} (-)^{ag_1} (-)^{ag_2}
\theta\left[^{a}_{b}\right]^2 \theta\left[^{a}_{b+1}\right]\theta\left[^{a}_{b-1}\right] = 0 $.} 
has an $N=4$ structure and thus all choices for the left R-symmetry operators are equivalent. The
quantum numbers of the
Euclidean time cycle and the internal $X_5$-cycle are coupled
to the spin structures $(a,b)$ in the same way. After performing the Jacobi theta-function identity
the partition function becomes:
$$\!\!\!\!\!\!\!\!\!\!\!\!\!\!\!\!\!\!\!\!\!\!\!\!\!\!\!\!\!\!\!\!\!\!\!\!\!\!\!\!Z_{\rm untwist} =~-{1\over 2}{V_5 \over (2\pi)^5}\oint_F {d\tau d\bar{\tau}\over ~4{\rm Im}\tau^{7
\over 2}~}\sum_{(n_1,m_1),(h_1,g_1)} \sum_{(n_2,m_2),(h_2,g_2)}(-)^{g_1+g_2+h_1+h_2}~ $$
 \be
~~~~~~~~~~~~~~~~~~~~~e^{-\pi R_0^2{|2m_1+g_1+(2n_1+h_1)\tau|^2\over {\rm Im} \tau}}~e^{-\pi
R_5^2{|2m_2+g_2+(2n_2+h_2)\tau|^2\over {\rm Im} \tau}}~\theta\left[^{1+h_1+h_2}_{1+g_1+g_2}
\right]^4~ {\Gamma_{(5,21)} \over \eta(\tau)^{12}~ {\bar\eta}(\bar\tau)^{24}}~.\ee 
The factor of
$1/2$ is due to the $\Z_2$ orbifolding of the $N=4$ theory.

\noindent Proceeding as in the simpler examples before and neglecting the $h_1=1$ and $h_2=1$ sectors for
large $R_0,\, R_5$, the non-zero contributions to the partition function occur when $g_1+g_2=1$.
Assuming also that all other moduli are of order unity, the only non-exponentially suppressed
contributions come from the zero mass left- and right-levels. We obtain 
\be 
Z_{\rm untwist}
=2^3D_0~{V_5 \over (2\pi)^5}~\sum_{g_1,g_2}{\left(1-(-)^{g_1+g_2}\right)\over 2}~ \int_0^{\infty
}{dt \over 2t^{7 \over 2}~}\sum_{m_1,m_2}~e^{-\pi R_0^2{(2m_1+g_1)^2\over t}-\pi
R_5^2{(2m_2+g_2)^2\over t}},
\label{partitionmixed}  
\ee which after the change of variables $t=\pi
\left(R_0^2 (2m_1 +g_1)^2~+~R_5^2 (2m_2 +g_2)^2\right)~x$ gives
$$
Z_{\rm untwist} = {4D_0~\Gamma\left({5\over 2}\right)\over \pi^{5\over 2}}~{V_5\over (2\pi)^5}~
\sum_{m_1,m_2}~{1 \over ( R_0^2(2m_1+1)^2+ R_5^2(2m_2)^2)^{5\over 2}}
$$
\be
 ~~~~~~~~~~~+{4D_0~\Gamma\left({5\over 2}\right)\over \pi^{5 \over 2}}~{V_5 \over
(2\pi)^5}~\sum_{m_1,m_2}~{1 \over ( R_0^2(2m_1)^2+ R_5^2(2m_2+1)^2)^{5\over 2}}.
 \ee
 This expression is symmetric under the $R_0\leftrightarrow R_5$ exchange. This is suggestive of a
temperature/gravitino mass, $T/M$, duality. This duality will be broken when the supersymmetry
breaking arises due to the coupling to a different $Q_R$ charge than $Q_F$.

\noindent To obtain the effective four dimensional pressure, we must factorize out the space-time
volume $V_4$. To this extent it is convenient to define the complex structure-like ratio
\be
u={R_0
\over R_5}={M \over T},
\ee
and re-write the partition function in the following way
$$
Z_{\rm untwist} = {4D_0~\Gamma\left({5\over 2}\right)\over \pi^{5 \over 2}}~{V_4 \over (2\pi
R_0)^4}~ \sum_{m_1,m_2}~{u^4 \over | (2m_1+1)iu+ 2 m_2|^{5}}
$$
\be 
 ~~~~~~~~~~~+{4D_0~\Gamma\left({5\over 2}\right)\over \pi^{5 \over 2}}~{V_4 \over (2\pi
R_5)^4}~\sum_{m_1,m_2}~{1 \over |2m_1 iu+ (2m_2+1)|^{5}}. 
\label{partitionmixed2} 
\ee Define the
function \be f(u)\equiv \sum_{m_1,m_2}~{u^4 \over | (2m_1+1)iu+ 2 m_2|^{5}},\ee which we can
express in terms of Eisenstein functions of order $5/2$: \be f(u) = u^{3\over 2}\left(\frac 1{2^{5
\over 2}}E_{5/2}\left({iu\over 2}\right) - \frac 1{2^{5}}E_{5/2}(iu)\right),\ee where \be E_k(U) =
\sum_{(m,n)\neq(0,0)}\left(\frac {{\rm Im}\ U}{|m+nU|^2}\right)^k. \ee Then the pressure in the
Einstein frame can be written as \be P_{\rm untwist} = ~C_T^{\rm unt}~{T^4}~ f(u)+C_V^{\rm unt}~
M^4~ { f(1/u) \over u} \label{pressuremixed}, \ee where \be C_T^{\rm unt}=C_V^{\rm unt}=n_{\rm
unt}^*~{\Gamma\left({5\over 2}\right)\over \pi^{5 \over 2}}.\ee Here $n_{\rm unt}^*=4D_0$ is the
number of massless boson/fermion pairs in the untwisted sector. It is smaller by a factor of $1/2$
from the corresponding number in the $N=4$ case due to the $\Z_2$-orbifolding. In this particular
model the coefficients $C_T^{\rm unt}$ and $C_V^{\rm unt}$ are equal due to the underlying
gravitino mass/temperature duality. For fixed $u$ the first term stands for the thermal
contribution to the pressure while the second term stands for minus the effective potential.

\noindent We note that the coefficient $C_T^{\rm unt}$ is fixed and positive as it is determined by the
number of all massless boson/fermion pairs in the untwisted sector of the initially supersymmetric
theory: $n_{\rm unt}^*=4D_0$. In general, the coefficient $C_V^{\rm unt}$  will depend on the
precise way supersymmetry is broken. As we have demonstrated in the previous section, it can take
both negative and positive values, depending on how the Susy-breaking operator couples to the right
movers: $C_V^{\rm unt} \sim {\tilde n}_{\rm unt}^*$. Thus in general the temperature/gravitino mass
scale duality will be broken.

\noindent Let us discuss the large $u$ limit, which can be obtained by taking $R_5$ to be small (but still
parametrically larger as compared to the string scale), while taking $R_0$ to be much larger. In
this limit we expect to find a four dimensional system at finite temperature, for which only
massless bosonic degrees of freedom are thermally excited. {\it All} fermions attain a mass from
the Scherk-Schwarz boundary conditions along the $X_5$ cycle, and this mass is much bigger than the
temperature for large $u$. Therefore they can be integrated out giving a temperature independent
contribution to the pressure of order $1/R_5^4$. Setting $\tilde{u}=1/u$, we have in the limit
$\tilde{u}\to 0$ ($u \to \infty$): \be f(u)=f(1/\tilde u)\to\sum_{m_1}\int_{-\infty}^{\infty}{dx
\over ((2m_1+1)^2 +4x^2)^{5 \over 2}}= {2\over 3}\sum_{m_1}{1 \over (2m_1+1)^4}={1\over
3}\times{\pi^4\over 24}, \ee and
$$
{f(1/u) \over u}\to \sum_{m_1}{1 \over |2m_1+1|^5}+{1\over 2^5~u^4}\sum_{m_2\ne
0}\int_{-\infty}^{\infty}{dx \over (m_2^2 +x^2)^{5 \over 2}}
$$
$$
=2\left(1- 2^{-{5}}\right)~\zeta(5)+{1\over 24~u^4}\sum_{m_2\ne 0}{1\over m_2^4}
$$
\be =2\left(1-2^{-{5}}\right)~\zeta(5)+{1\over 12~u^4}\times{\pi^4\over 90}. \ee Using
these results, we find \be P_{\rm untwist}={1 \over 3}~{2^3 D_0\over 2}~{\pi^2 T^4 \over
30}+2\left(1- 2^{-{5}}\right)~{\zeta(5)\Gamma\left({5\over 2}\right)\over \pi^{5 \over
2}}~(4 D_0)~M^4 \ee for the first two leading terms for small $\tilde u=1/u$. We have used the
relation $\Gamma({5 \over 2})=3\sqrt{\pi}/4$. The first contribution arises from the thermally
excited massless bosons. As compared to Eq.  (\ref{purethermalfield}) with $n_B=2^3D_0$ and
$n_F=0$, it is off by a factor of $1/2$ due to the $\Z_2$-orbifolding. Similarly, the second term is
off by a factor of $1/2$ as compared to Eq.  (\ref{VN4zeroT}) due to the orbifolding. For large
$u$, the potential term is dominant. For generic values of $u$ both fermions and bosons contribute
to the thermal piece as in Eq.  (\ref{pressuremixed}), with the contribution depending on the
number of massless states at zero temperature and before the breaking of supersymmetry. Finally for
small $u$, the system is essentially a five dimensional purely thermal system.

\subsection{Twisted sector, $H=1$}

The contributions of the twisted sectors in the large $R_0, R_5$ limit depend on the number of the
massless twisted states before the supersymmetry breaking, and can be determined in a similar way as
before. However, there is a class of models where the $\Z_2$-orbifolding acts freely, without any
fixed points, and therefore there are no massless states in the twisted sectors. For this class of
models, the whole contribution to the one-loop partition function, in the large $R_0, R_5$ limit,
is that of the massless untwisted sector states we have already determined. One example with this
property is when the $\Z_2$-twists $(H,G)$ are accompanied with a shift of the $\Gamma_{(1,1)}(R_6)$
sub-lattice. This operation leads to the modification of $Z_{(2,10)}$ in Eq. 
(\ref{N2partition}), where we also set $(n_0, n_t)=(8, 8)$, as follows:
\be
Z_{(2,10)}\left[^0_0\right]~~\longrightarrow~~Z_{(2,10)}\left[^{H}_{G}\right]=
{\Gamma_{(1,1)}(R_5)~ \Gamma_{(1,1)}(R_6)\left[^H_G \right]~ \over \eta(\tau)^2~ {\bar\eta}({\bar
\tau})^{2} }~{1 \over 2}\sum_{\gamma,\delta} {{\bar\theta} \left[^{\gamma}_{\delta}\right]^{8}
\over {\bar\eta}({\bar \tau})^{8} }.
\ee
If $R_6$ is sufficiently large, the coupling of the
$(H,G)$-shift of the lattice to the twisted partition function $Z_{(4,12)}\left[^H_G \right]$
ensures the absence of massless states in the twisted sector.

\noindent In other situations, there are massless states in the twisted sector. Before the supersymmetry
breaking, the number of massless bosons is equal to the number of massless fermions with a
multiplicity $n^*_{\rm twist}$. Proceeding as in the untwisted sector, and neglecting the
$h_1,h_2=1$ sectors, one finds that there is only a non zero contribution when $g_1+g_2=1$. The
relative sign of the thermal part $\sim T^4$ and the supersymmetry breaking part $\sim M^4$ depends
on the choice of the operators $Q_F$ and $Q_R$. When there is no coupling to the right-moving gauge
charges we obtain:
 \be
  P_{\rm twist} = ~C_T^{\rm twist}~{T^4}~ f(u)+C_V^{\rm twist}~ M^4~ { f(1/u)
\over u}, \label{twistedpressure1}
\ee where
 $$C_T^{\rm
twist}=n^*_{\rm twist}~{\Gamma\left({5\over 2}\right)\over \pi^{5 \over 2}}
$$
$$C_V^{\rm
twist}=\epsilon~n^*_{\rm twist}~{\Gamma\left({5\over 2}\right)\over \pi^{5 \over 2}}
$$
\be
\epsilon =(-)^{(Q_R-Q_F)}. \label{twistedpressure2}
\ee
Here, we have for the coefficient
$\epsilon$:

$\bullet$ $\epsilon =1$ when $Q_R=Q_F$.

$\bullet$ $\epsilon =-1$ when $Q_R\neq Q_F$.

\noindent In the later case, $(-)^{(Q_R-Q_F)}=(-)^{H}=-1$, see Eq. (\ref{Rshifted}), and $H=1$ in the
twisted sector. The change of sign indicates that it is the bosons that are becoming massive
because of the supersymmetry breaking. This is related to a
mechanism for moduli stabilization induced by geometrical fluxes \cite{GeometricalFluxes}.

\noindent Adding the contributions of the untwisted and twisted sectors together we obtain for the pressure
\be P=~C_T~{T^4}~ f(u)+C_V~ M^4~{f(1/u)\over u} , \ee where $C_T=C_T^{\rm unt}+C_T^{\rm twist}$ and
likewise for $C_V$. The sign of the thermal contribution is always positive,
\be C_T=
{n^*_T~\Gamma\left({5\over 2}\right)\over \pi^{5 \over 2}},
\ee
$n_T^*=n^*_{\rm unt}+n^*_{\rm
twist}$. The coefficient multiplying the supersymmetry breaking part is given by
\be
C_V=
{n^*_V~\Gamma\left({5\over 2}\right)\over \pi^{5 \over 2}}
\ee with $n_V^*=n^*_{\rm
unt}+\epsilon~n^*_{\rm twist}$. In general $n_V^*$ can be positive or negative depending on the
model.

\subsection{An explicit example}
\label{E7SU2} As an example we consider the $E_8\times E_8$ heterotic string on a $T^4/\Z_2$
orbifold, whose initially supersymmetric partition function is obtained by setting
\be
Z_{(2,10)}=
{\Gamma_{(1,1)}(R_5)~ \Gamma_{(1,1)}(R_6)~ \over \eta(\tau)^2~ {\bar\eta}({\bar \tau})^{2} }~{1
\over 2}\sum_{\gamma,\delta} {{\bar\theta} \left[^{\gamma}_{\delta}\right]^{8} \over
{\bar\eta}({\bar \tau})^{8} },\ee and \be Z_{(4,12)}\left[^H_G\right]=
{\Gamma_{(4,4)}\left[^H_G\right]~ \over \eta(\tau)^4~ {\bar\eta}({\bar \tau})^{4} }~{1 \over
2}\sum_{\gamma^{\prime},\delta^{\prime}} {{\bar\theta}
\left[^{\gamma^{\prime}}_{\delta^{\prime}}\right]^{6}{\bar\theta}
\left[^{\gamma^{\prime}+H}_{\delta^{\prime}+G}\right]{\bar\theta}\left[^{\gamma^{\prime}-H}_{\delta^{\prime}-G}\right]
\over {\bar\eta}({\bar \tau})^{8} }
\ee in Eq.  (\ref{N2partition}). We shall use an R-symmetry
operator associated to one of the twisted complex planes for breaking the supersymmetry, replacing
the $\Gamma_{(1,1)}(R_5)$ lattice as in Eq. (\ref{Rshifted}). In the twisted sectors,
$(H,G)\neq (0,0)$, the internal $\Gamma_{(4,4)}$ shifted lattice is given by
\be
\Gamma_{(4,4)}\left[^H_G\right]= {2^4\eta(\tau)^6~ {\bar\eta}({\bar \tau})^{6}
\over\theta\left[^{1+H}_{1+G}\right]^2{\bar\theta}\left[^{1+H}_{1+G}\right]^2}.
\ee
The orbifolding
breaks the $E_8\times E_8$ gauge group to $E_8\times E_7\times SU(2)$. Under $E_8 \to  E_7\times
SU(2)$, the 248-dimensional adjoint representation of $E_8$ decomposes as \be ({\bf 1},{\bf 3})\oplus ({\bf 56},{\bf 2})\oplus ({\bf 133},{\bf 1}).
\ee

\noindent The untwisted sector contains $2^3\times 504$ massless states giving the value $n^*_{\rm
unt}=4\times 504$ for the total number of boson/fermion pairs. These numbers arise as follows. In
terms of world-sheet left/right movers the number of bosonic degrees of freedom is given by
$$
\!\!\!\!\!\!\!\!n^*_{\rm unt}=4\times 504=[4]_{\psi^{2,3,5,6}}\times\left([4]_{X^{2,3,5,6}}+[248]_{E_8}+
[133]_{E_7}+[3]_{SU(2)}\right)+
$$
\be
~~~~~\,[4]_{\psi^{7,8,9,10}}\times\left([4]_{X^{7,8,9,10}}+ [2]_{SU(2)}\times[56]_{E_7}\right).
\ee
The first line gives the bosonic content of a $d=6$ supergravity multiplet, a tensor multiplet and
a vector multiplet in the adjoint of the $E_8\times E_7\times SU(2)$ gauge group. The second line
gives the bosonic content of four uncharged and one charged hyper-multiplets. The number of
fermionic degrees of freedom follows by supersymmetry.  At finite temperature and when Susy is
broken, the contribution of the massless untwisted states is determined as before (see Eq. 
(\ref{partitionmixed2})).

\noindent Next we analyze the contribution of states in the twisted sectors, $H=1$. For large $R_0, R_5$, we
may neglect the $h_1=1$ and $h_2=1$ sectors. Also for $R_6$ of order unity we may set
$\Gamma_{(1,1)}(R_6)\cong 1$. Setting $H=1$, the partition function becomes in this limit
$$
\!\!\!\!\!\!\!\!Z_{\rm twisted}={V_5 \over (2\pi)^5}\int_{||} {d\tau d\bar{\tau}\over 4{\rm Im}\tau^{7 \over 2}~}~
{1 \over\eta(\tau)^{6}\bar\eta(\bar\tau)^{18}} ~\sum_{(m_1,g_1),(m_2,g_2)}e^{-\pi
R_0^2{(2m_1+g_1)^2\over {\rm Im} \tau}} e^{-\pi R_5^2{(2m_2+g_2)^2\over {\rm Im} \tau}}
$$
$$
~~~~~~\times {1\over 4}\sum_{(a,b,G)} (-)^{a+b+ab}~(-)^{a g_1}~(-)^{(a+1)g_2}
~~{\theta\left[^a_b\right]^2 \theta\left[^{a+1}_{b+G}\right]^2}
$$
\be
~~~~~~~~~~~~~~~~~~~~~~~~~~~~~~~\times {2^4 \over\theta\left[^{0}_{1+G}\right]^2{\bar\theta}\left[^{0}_{1+G}\right]^2}~~{1
\over 2}\sum_{\gamma,\delta} {\bar\theta} \left[^{\gamma}_{\delta}\right]^{8}~~{1 \over
2}\sum_{\gamma^{\prime},\delta^{\prime}} {\bar\theta}
\left[^{\gamma^{\prime}}_{\delta^{\prime}}\right]^{6}{\bar\theta}
\left[^{\gamma^{\prime}+1}_{\delta^{\prime}+G}\right]{\bar\theta}\left[^{\gamma^{\prime}-1}_{\delta^{\prime}-G}\right]
. \ee
The non-vanishing contributions arise when $g_1+g_2=1$. Non-exponentially suppressed
contributions arise only at the zero mass level. To obtain them, we expand the integrand in powers
of $q=e^{2\pi i\tau}$.

\noindent When $g_1+g_2=1$, we have for the left movers
$$
\!\!\!\!\!\!\!\!\!\!\!\!\!\!\!\!\!\!\!\!\!\!\!\!\!\!\!\!\!\!\!\!\!\!\!\!\!\!\!\!{1 \over \eta(\tau)^{6}~\theta\left[^{0}_{1+G}\right]^2}~~  {1\over 2}\sum_{(a,b)}
(-)^{a+b+ab}~(-)^{a g_1}~(-)^{(a+1)g_2} ~~{\theta\left[^a_b\right]^2
\theta\left[^{a+1}_{b+G}\right]^2}
$$
$$
~~~~~~~~~~~~~~={1 \over \eta(\tau)^{6}~\theta\left[^{0}_{1+G}\right]^2}~~{1\over 2}\sum_{(a,b)}(-)^{g_2}
(-)^{b+ab} ~~{\theta\left[^a_b\right]^2 \theta\left[^{a+1}_{b+G}\right]^2}
$$
$$
~~~~~~~~~~~~~~~~~~~={(-)^{g_2}\theta_2^2\theta_3^2 \over \eta^6 ~\theta_4^2} ~~
{\rm for}~~  G=0~~~~Ê{\rm or}~~~~ {(-)^{g_2}\theta_2^2\theta_4^2 \over \eta^6~\theta_3^2} ~~
{\rm for} ~~ G=1
$$
\be \!\!\!\!\!\!\!\!\!\!\!\!\!\!\!\!\!\!\!\!\!\!\!\!\!\!\!\!\!\!\!\!\!\!\!\!\!\!\!\!= 4(-)^{g_2}(1+{\cal O}(q^{1/2})) \, .
 \ee
 For the right movers we have
$$
\!\!\!\!\!\!\!\!\!\!\!\!\!\!\!\!\!\!\!\!\!\!\!\!\!\!\!\!\!\!\!\!\!\!\!\!\!\!\!\!\!\!\!\!\!\!\!\!\!\!\!\!\!\!\!\!\!\!\!\!\!\!\!\!\!\!\!\!\!\!\!\!\!\!\!\!\!\!\!\!{ 1 \over \bar\eta(\bar\tau)^{18}~{\bar\theta}\left[^{0}_{1+G}\right]^2}~~{1 \over
2}\sum_{\gamma,\delta} {\bar\theta} \left[^{\gamma}_{\delta}\right]^{8}~~ {1 \over
2}\sum_{\gamma^{\prime},\delta^{\prime}} {\bar\theta}
\left[^{\gamma^{\prime}}_{\delta^{\prime}}\right]^{6}{\bar\theta}
\left[^{\gamma^{\prime}+1}_{\delta^{\prime}+G}\right]{\bar\theta}
\left[^{\gamma^{\prime}-1}_{\delta^{\prime}-G}\right]
$$
$$
={1 \over \bar q^{3/ 4}}\left(1+(-)^G\,  4\ \bar q^{1/2}+ \dots\right)\left(1 + 240\ \bar q+\dots\right)
\left[\bar q^{1/4}\left((-)^G\,  {2} + 56\ \bar q^{1/2}+\dots\right)\right]
$$
\be \!\!\!\!\!\!\! \!\!\!\!\!\!\!\!\!\!\!\!\!\!\!\!\!\!\!\!\!\!\!\!\!\!\!\!\!\!\!\!\!\!\!\!\!\!\!\!\!\!\!\!\!\!\!\!\!\!\!\!\!\!\!\!\!\!\!\!\!\!\!\!\!\!\!\!\!\!\!\!\!\!\!\!\!\!\!\!={1 \over \bar q^{1/ 2}}\left[(-)^G\,  2+8\ \bar q^{1/2} + 56\ \bar q^{1/2} +{\cal O} (\bar q)\right] \, . \ee When we add the contributions of the $(H,G)=(1,0)$ and $(H,G)=(1,1)$ twisted
sectors together, we find that the lowest right-mass level is at zero mass as it is the case for
the lowest left-mass level.

\noindent Using these results we obtain the contribution to the partition function of
the massless twisted states:
$$
 \!\!\!\!\!\!\!Z_{\rm twist}=2^6 (56+8){V_5 \over (2\pi)^5} \sum_{g_1,g_2}{(-)^{g_2}-(-)^{g_1}\over 2}
\int_0^{\infty} {dt\over 2t^{7 \over 2}} \sum_{(m_1,m_2)}e^{-\pi R_0^2{(2m_1+g_1)^2\over {\rm Im}
\tau}} e^{-\pi R_5^2{(2m_2+g_2)^2\over {\rm Im} \tau}}
$$
\be \!\!\!\!\!\!\! \!\!\!\!\!\!\! \!\!\!\!\!\!\! \!\!\!\!\!\!\! \!\!\!\!\!\!\! \!\!\!\!\!\!\! \!\!\!\! ={2^5 (56+8)~\Gamma\left({5\over 2}\right)\over \pi^{5 \over 2}}~\left({V_4 \over (2\pi
R_0)^4}~ f(u)-~{V_4 \over
(2\pi R_5)^4}{f(1/u) \over u}\right). \ee The contribution to the pressure is as in Eqs
(\ref{twistedpressure1}), (\ref{twistedpressure2}) with $n^*_{\rm twist}=2^5( 56+8)$, the
number of massless boson/fermion pairs in the twisted sector, and $\epsilon=-1$. The massless
bosonic content consists of $32$ scalars in the $(56,1)$ representation of $E_7 \times SU(2)$ and
$128$ scalars in the representation $(1,2)$. The number of massless fermionic degrees of freedom follows by
supersymmetry.

\noindent Adding the contributions of the untwisted and twisted sectors together, we obtain
$$
n_T^*=4\times 504+16\times128=4064
$$
\be n_V^*=4\times 504-16\times128=-32. \ee In addition, we have the choice with $\epsilon=1$ in
Eq. (\ref{twistedpressure2}), giving $n_T^*=n_V^*=4064$. We can also change $n_V^*$ by
considering Susy-breaking operators involving the right-moving gauge charges as in Eq. 
(\ref{Qgauge}).

\subsection{Small mass scales from Wilson line deformations}
A generic supersymmetric heterotic background may contain in its spectrum massive supermultiplets
whose mass is obtained by switching on non-trivial continuous Wilson-lines \cite{Kiritsis} --\cite{Kiritsis:2007zz}.
This is a stringy
realization of the Higgs mechanism, breaking the initial gauge group $G$ to a smaller one
spontaneously. This statement is not absolutely correct for discrete Wilson lines corresponding to
extended symmetry points where the gauge symmetry may enhance or even get modified.

\noindent For our purposes, we restrict to arbitrary and small Wilson line deformations starting from a given
supersymmetric background where $R_I,\, I=6, 7, \dots, 10$ are of the order the string scale. This
restriction ensures that the contributions to the thermal partition function of the momentum and
winding states in these five internal directions will be exponentially suppressed in the limit
where $R_0$ and $R_5$ are large.

\noindent A systematic study of the effects of Wilson lines can be found in \cite{Kiritsis} --\cite{Kiritsis:2007zz}. In the zero
winding sector, a Wilson line just modifies the Kaluza-Klein momenta, and the corresponding
Kaluza-Klein mass becomes
\be {m_I^2\over R_I^2}~~\longrightarrow~~  {(m_I+~y_I^a~Q_a)^2\over
R_I^2},
\ee
where $Q_a$ is the charge operator associated to the Wilson-line $y_I^a$. We distinguish
three different situations according to the direction $I$:
\\
i) $I=5$, where $R_5$ is larger than the string scale.\\
ii) $I=6$, with $R_6$  of the order of the string scale (with $N=4$ and $N=2$ initial supersymmetry).\\
iii) $I=6,\dots,10$,  with $R_I$  of the order of the string scale (with $N=4$ initial susy only).
 
\noindent In the first case, $I=5$, after a Poisson re-summation, the net modification to the partition
function is obtained by the following replacement in Eq.  (\ref{partitionmixed2}):
$$ \int_0^{\infty }{dt \over t^{7 \over 2}~}\sum_{m_1,m_2}~e^{-\pi
R_0^2{(2m_1+g_1)^2\over t}-\pi R_5^2{(2m_2+g_2)^2\over t}}~\longrightarrow~~$$ \be ~ \int_0^{\infty
}{dt \over t^{7 \over 2}~}\sum_{m_1,m_2}~e^{-\pi R_0^2{(2m_1+g_1)^2\over t}-\pi
R_5^2{(2m_2+g_2)^2\over t}}~\left[ e^{2i\pi(2m_2+g_2)~y_5^a Q_a}\right].\ee
The term in the brackets can be replaced with \be \cos\left(2\pi (2m_2+g_2)~y_5^a
Q_a\right)=1-2\sin^2\left(\pi (2m_2+g_2)~y_5^a Q_a\right). \ee

\noindent In the ii) and iii) cases, we can set the momentum and winding numbers to zero,
$m_I=n_I=0$, so that the extra modification in the partition function is the insertion of the term:
\be \left[e^{-\pi t \sum_I \left({y_I^a Q_a\over R_I}\right)^2}\right]~\simeq ~\left[1~{-\pi t
\sum_I \left({y_I^a Q_a \over R_I}\right)^2}\right].\ee

\noindent Incorporating the effects of the Wilson lines up to quadratic order, we get for the overall
pressure:
$$
 \!\!\!\!\!\!\!P =T^4~{\Gamma\left({5\over 2}\right)\over
\pi^{5\over2}}\sum_{m_1,m_2}{u^4~\left(~n^*_T-2\sum_s\sin^2(2 \pi
m_2y_5^aQ_a^s)~\right)\over|(2m_1+1)iu+2m_2|^5}
$$
$$
~~~~~-~T^2~{\Gamma\left({3\over 2}\right)\over\pi^{3\over2}}\sum_{m_1,m_2}
{u^2\left(~M_T^2-2\sum_sM_s^2\sin^2(2 \pi m_2y_5^aQ_a^s)~\right) \over|(2m_1+1)iu+2m_2|^3}
$$
$$
~~~~~~~~~~~~+~M^4~{\Gamma\left({5\over 2}\right)\over\pi^{5\over2}}~\sum_{m_1,m_2} {~n^*_V-2\sum_s {\rm
sign}(s) \sin^2\left((2m_2+1)\pi y_5^aQ_a^s\right)~\over|2m_1iu+(2m_2+1)|^5}
$$
\be \label{P} ~~~~~~~~~~~~~~~~~~~~~~~~-~M^2~{\Gamma\left({3\over 2}\right)\over\pi^{3\over2}}\sum_{m_1,m_2}
{M_V^{(2)}-2\sum_s {\rm sign}(s)M_s^2 \sin^2\left((2m_2+1)\pi y_5^aQ_a^s\right) \over|2m_1 iu
+(2m_2+1)|^3}. \ee In this expression, we have defined 
\be
M^2_s= {1\over 4\pi }\sum_{I} {(y_I^aQ_a^s)^2\over R_I^2}
\ee
 for the pair of boson/fermion
states $s$ and also introduced \be M_T^2=\sum_s M_s^2\; , \quad M_V^{(2)}=\sum_s {\rm
sign}(s)M_s^2\, , \ee where ${\rm sign}(s)$ indicates whether the state $s$ contributes positively
or negatively to $n_V^*$ and $M_V^{(2)}$, both being possibly negative.

\noindent The following comments are in order:\\
\indent $\bullet$ In the above expression, $y_5^aQ_a^s$ summarizes  the
contribution of the $R_5$-Wilson line in the term corresponding to the pair of boson/fermion states
$s$. It does not introduce a new scale.\\
\indent $\bullet$ The $M_s$'s introduce new mass scales in the theory, qualitatively different than $T$ and $M$.
The masses $M_s$ are supersymmetric mass scales rather than susy-breaking scales like $T$ and
$M$. One must keep in mind that after minimization with respect to $y^a_I$, 
physical scales like $M_s$ are always proportional to an (infrared) renormalization group invariant scale like, for instance, 
$\Lambda_{\rm {QCD}}$ or the transmutation scale $Q_0$,
appearing in the electroweak gauge symmetry radiative breaking 
(of supersymmetric theories in the presence of soft breaking terms) \cite{noScale}\cite{KounnasPavel}.   \\
 \indent $\bullet$ The first two terms (which arise from the $(g_1,g_2)=(1,0)$ sector) can be identified in
the effective field theory as the thermal contribution to the pressure, $P_{\rm thermal}$. Again
the number $n^*_T$ is always positive being the number of the massless boson/fermion pairs in the
initially supersymmetric background. This purely thermal piece is always positive.\\
\indent $\bullet$ The two
last terms can be identified as minus the effective potential $-{\cal V}_{\rm eff}$. This is
naturally regularized in the infrared by the temperature scale $T$.
This infrared regularization differs from that considered in \cite{StringInfrared}, and used in \cite{KP},
which is valid at zero temperature.\\
\indent $\bullet$ The number $n^*_V$ can be either positive or negative depending on the
way supersymmetry is broken. This shows that the sign of the one loop effective potential depends
on the way supersymmetry is broken, as it can be seen in supergravity by utilizing super-trace
arguments \cite{FerraraKounnasZwirner}\cite{KounnasPavel}.

\subsection{Scaling properties of the thermal effective potential}
The final expression for $P$ contains various mass scales: the two supersymmetry breaking scales
which are the temperature $T$ and the gravitino mass scale $M$, as well as the supersymmetric
masses $M_s$ which are generated by the Wilson-lines in the directions $6,7,8,9,10$. The first
identity follows immediately from the definition of $P$:
\be
\left(T{\partial \over \partial T}
+M{\partial \over
\partial M}+\sum_s M_s{\partial \over \partial M_s}\right)~P=4P
\ee
which can be best seen by writing $P$ as
\be
\label{Ifunctions}
P\equiv T^4~p_4(u)~
+~T^2~p_2(u),~~~~u={M\over T},
\ee
where
\be
\label{FG}
p_4= {\Gamma\left({5\over 2}\right)\over\pi^{5\over2}}
\left( F(u,y_5^a)+\tF(u,y_5^a)\right)\; ,\quad p_2= -{\Gamma\left({3\over
2}\right)\over\pi^{3\over2}} \left( G(u,y_5^a)+\phantom{\tF}\!\!\!\tG(u,y_5^a)\right)\, ,
\ee
and
using the definitions
$$
\!\!\!\!\!\!\!\!\!\!\!\!\!\!\!\!F(u,y_5^a)=\sum_{m_1,m_2, s} {u^4\cos(4\pi  m_2
y_5^aQ_a^s)\over[(2m_1+1)^2u^2+4m_2^2]^{5/2}},
$$
\be
\label{f}
~~~\tF(u,y_5^a)=\sum_{m_1,m_2,s} {u^4{\rm sign}(s)\cos
(2\pi(2m_2+1)y_5^aQ^s_a)\over[4m_1^2u^2+(2m_2+1)^2]^{5/2}}; \ee
$$
\!\!\!\!\!\!\!\!\!\!\!\!\!\!\!G(u,y_5^a)=\sum_{m_1,m_2,s} {u^2M_s^2\cos(4 \pi m_2y_5^aQ^s_a)\over[(2m_1+1)^2u^2+4m_2^2]^{3/2}},
$$
\be
\label{g}
~~~~~~~~\tG(u,y_5^a)=\sum_{m_1,m_2,s} {u^2{\rm sign}(s)M_s^2\cos(2\pi
(2m_2+1)y_5^aQ^s_a)\over[4m^2_1u^2+(2m_2+1)^2]^{3/2}} . \ee

\noindent Using standard thermodynamics
identities, we can obtain the energy density $\rho$: \be \label{defrho} \rho \equiv T{\partial P \over \partial
T}-P = T^4~r_4(u)~+~T^2~r_2(u) \ee where \be \label{r42} r_4=3p_4-u p_4^{\prime}\; , \quad
r_2=p_2-u p_2^{\prime} \ee and the primes stand for derivatives with respect to $u$. In the sequel,
we allow the Susy-breaking scales $T$ and $M$ to vary with time while fixing the supersymmetric
masses $M_s$, and investigate the back-reaction to the initially flat metric and moduli fields.


\section{Gravitational equations and critical solution}
\noindent We assume that the back-reacted space-time metric is
homogeneous and isotropic,
\be
ds^2=-dt^2+a(t)^2~d\Omega_k^2,~~~~H=\left({\dot{a}\over a}\right),
\ee where $\Omega_k$ denotes the three dimensional space with
constant curvature $k$ and $H$ is the Hubble parameter.

\noindent From the fact that $-P$ plays the role of the effective
potential
and the relation between the gravitino mass scale $M$ and the no-scale modulus $\Phi$,
$$
M=e^{\alpha\Phi},~~~~\alpha=\sqrt{3\over2},
$$
we obtain the field equation for $\Phi$: \be
\ddot{\Phi}+3H\dot{\Phi}={\partial P\over \partial \Phi}=\alpha
u\left({\partial P \over \partial u}\right)_T =-\alpha\left(
T^4(r_4-3p_4) +T^2(r_2-p_2)\right).
\label{Phi}
\ee
We have made use of Eq. 
(\ref{r42}).

\noindent For other flat moduli $\varphi_i$, with $\Phi$ independent
kinetic terms, the equation of motion is straightforward to solve,
\be
 \ddot \varphi_i + 3H\dot\varphi_i=0\quad \Longrightarrow \quad {1\over
2}\dot \varphi_i^2={c_i^2\over a^6}\, ,
\label{phiI}
\ee
where the $c_i$'s are
integration constants.

\noindent Knowing the thermal effective potential $-P$, the
energy density $\rho$ as well as the field equation for the modulus
$\Phi$, we can derive the (one-loop) corrected space-time metric by
solving the gravitational field equations. These are the
Friedmann-Hubble equation,
\be
\label{Hubble} 3H^2={1\over
2}\dot\Phi^2+{1\over 2}\sum_i \dot\varphi_i^2+\rho-{ 3 k\over a^2}, \ee
and the equation that follows from varying with respect to the
spatial components of the metric: \be 2\dot H +3H^2 =-{k\over
a^2}-P-{1\over2} \dot\Phi^2-{1\over 2}\sum_i \dot\varphi_i^2\, .
\label{vara} \ee For our purposes, it will be useful to replace Eq. 
(\ref{vara}) by the linear sum of Eqs (\ref{Hubble}) and
(\ref{vara}), so that the kinetic terms of $\Phi$ and $\varphi_i$ drop
out: 
\be 
\label{gravityeq} 
\dot H +3H^2 =-{2k\over a^2}+{1\over
2}(\rho-P)\, . 
\ee

\subsection{Critical solution}

The fundamental ingredients in our analysis are the scaling
properties of the thermal effective potential $-P=-T^4p_4-T^2p_2$.
These scaling properties suggest to search for a solution where all
varying mass scales of the system, $M(\Phi)$, $T$ and $1/ a$,
remain proportional during time evolution: \be \label{anzats}
e^{\alpha\Phi} \equiv M(\Phi)={1\over {\gamma}a}~~\Longrightarrow
~~H=-{\alpha} \dot{\Phi},~~~~ M(\Phi)\,= u \,T \, ,\ee with $\gamma$
and $u$ fixed in time. Our aim is to determine the constants
$\gamma$ and $u$.

\noindent On the trajectory (\ref{anzats}), the $\Phi$-equation is
given by
\be
\dot H + 3H^2 =\alpha^2\left((r_4-3p_4){M^4\over
u^4}+(r_2-p_2){M^2 \over u^2} \right) 
\label{PhiBare}
\ee 
and the gravity Eq. 
(\ref{gravityeq}) by 
\be \label{gravcri} \dot H +
3H^2=-2k\gamma^{2}M^2+ {1\over 2}(r_4-p_4){M^4\over u^4}  + {1\over
2}(r_2-p_2){M^2\over u^2} \, .
\ee
The compatibility of these two
equations requires an identification of the coefficients of the
monomials in $M$. The quartic terms give an equation for $u$, while
the quadratic terms determine the sign of the parameter $k$ and the
magnitude of $|k\gamma^2|$,
\be
\label{equ} r_4= {6\alpha^2-1\over
2\alpha^2-1}p_4\, ,~~~\left(r_4=~4p_4\, , ~~\mbox{{\rm \ie}}~~ p_4+up_4'=0\, ,~~{\rm
for}~~\alpha^2={3\over2}\right) , 
\ee 
\be \label{eqk}
-2k\gamma^{2}~=~{2\alpha^2-1 \over 2} ~(r_2-p_2)~{1\over
u^2}\, ,~~~\left( -2k\gamma^{2}={(r_2-p_2)\over u^2}\, ,~~~~{\rm
for}~~\alpha^2={3\over 2}\right).
\ee
Eq. (\ref{equ}) reminds
us of the equation of state for thermal radiation in
five dimensions. In the absence of Wilson lines, where $r_2=p_2=0$,
we have that $\rho=4P$, which is indeed the $5d$ state equation
for thermal radiation. 

\noindent
When non trivial Wilson lines are turned on, 
their equations of motion fix them to be proportional to the renormalization group invariant scale $Q_0$.  
$Q_0$ is supposed to be much smaller than $M$, $Q_0\ll M$, so that the $y^4$-terms can be consistently neglected. 
Within this approximation the relevant terms of the $y^a_I$-effective potential are quadratic in $y^a_I$ modulo multiplicative 
terms arising from wave-function renormalization. If we define
\be
p_2=\sum_{I,a}\left({y_I^a\over R_I}\right)^2p_{2I}^{a}(u)\, ,
\ee
we obtain, in a collectively qualitative way, the effective potential for each Wilson line (no sum over $I$ and $a$):
 \be
 -V^{a}_I=Z^{a}_I(y^a_I/R_I,Q_0)\,T^2 \,\left({y_I^a\over R_I}\right)^2  p^{a}_{2I}(u),~~~~~~ Z^{a}_I(y^a_I/R_I,Q_0)=A^{a}_I\left(1-\log~{(y_I^a/R_I)^2\over c^a_I Q_0^2}\right). 
 \ee
So we have that
 \be
 \label{Cau}
(y_I^a/R_I) { \partial\ \over  \partial (y_I^a/R_I)} V^{a}_I =0~~~~ \Longleftrightarrow ~~~~~ 
 \left({y_I^a\over R_I}\right)^2=c^a_I Q_0^2  ~,
 \ee
fixing the Wilson lines to be proportional to the transmutation scale $Q_0$. We stress at this point that the $c^a_I$ 
are positive dimensionless constants, which, in general, are functions of the dimensionless ratio  $u=M/T$. 
The $u$ dependence of the $c^a_I$ is due to threshold corrections induced in the large scale regime: $M,T\gg Q_0$. 
As we will show below, the $u$ dependence of the $c^a_I$ plays a crucial role in our analysis.      

\noindent
When all other equations of motion are satisfied, the gravity Eq. (\ref{gravityeq}) is equivalent to the total energy-momentum conservation:
\be
{ d \over dt}\left (\rho+ {1\over
2}\dot\Phi^2+{1\over 2}\sum_i \dot\varphi_i^2 \right)+3H\left(\rho+P+ \dot\Phi^2+\sum_i \dot\varphi_i^2 \right)=0\, .
\label{consEnergy} 
\ee
This imposes an extra constraint when Wilson lines are switched on. On the critical trajectory (\ref{anzats}), and by using (\ref{Cau}),
one obtains :
\be
r^R_2+p^R_2=0\, ,
\label{r2p2}
\ee 
where we have defined the renormalized quantities
\be
\label{p2R}
p_2^R=\sum_{I,a}Z^a_I\, \left({y_I^a\over R_I}\right)^2\, p_{2I}^a=\sum_{I,a}A^a_I\, c_I^a(u)Q_0^2\, p_{2I}^a(u)\; ,\quad r_2^R=p_2^R-up_2^{R\prime}\, ,
\ee
to appear as well in Eqs (\ref{PhiBare}), (\ref{gravcri}) and (\ref{eqk}).

\noindent Next, let us consider the Friedmann-Hubble Eq. 
(\ref{Hubble}) along the critical trajectory (\ref{anzats}). It  becomes
\be
 \label{Hubblenew}
\left({6\alpha^2-1\over 6\alpha^2}\right)~3H^2 =  -{3k\over a^2}+
\rho +{1\over 2}\sum_i {\dot \varphi_i}^2= -{3k\over a^2}+T^4 r_4+T^2r^R_2+
\sum_i {c_i^2\over a^6} \, .
 \ee
The dilatation factor in front of $3H^2$ can be absorbed
in the definition of $\hat k$, $c_r$ and $c_m$, once we take into
account Eqs (\ref{keff}) , (\ref{CReff}) and  (\ref{Cm}) below:
 \be
\label{cosmoeff} 3H^2= -{ 3\hat k\over a^2} +{c_r \over
a^4}+{c_m\over a^6}\, ,
\ee
where, for $\alpha^2=3/2$,
 \be 
 \label{keff} 
 \hat k=-{3\over
2\gamma^2 u^2} ~r^R_2~=~{3\over
2\gamma^2 u^2} ~p^R_2\, ,
\ee
\be
\label{CReff} 
c_r={9\over 8\gamma^{4} u^4}~r_4~= ~{9\over 8\gamma^{4} u^4}~4p_4\, , 
\ee 
and
 \be 
 \label{Cm} 
 c_m={9\over
8} \sum_i c_i^2\, .
 \ee 
Clearly, a necessary condition for
the effective curvature $\hk$ not to vanish is to have non trivial Wilson
lines, at least in the direction 6, and to ensure Eq. (\ref{r2p2}). 

\noindent We note that Eq. (\ref{cosmoeff}) also controls the
dynamics of a FRW universe, where space has constant curvature
$\hk$ and is
formally filled with a thermal bath of radiation (since the sign of
$c_r$ can a priori be positive as well as negative). There can be an
extra contribution, arising from the kinetic terms of some extra flat moduli,
that scales as $1/a^6$.

\noindent Finally let us address a seeming puzzle. We said that along the
critical trajectory and in the absence
of Wilson lines, our thermodynamic quantities satisfy $\rho=4P$. How
can this situation correspond to a $4d$ universe filled with thermal
radiation? The answer is that we must take into account the kinetic
energy density of the modulus field $\Phi$. When there are no Wilson
lines and the kinetic terms of the other moduli are switched off, the
Friedmann-Hubble equation gives ${\dot \Phi}^2 =(r_4T^4)/4$ along the critical
trajectory, and so the total energy density and pressure satisfy
$$
\rho_{\rm tot}=r_4T^4+ {1 \over 2}\dot\Phi^2={9r_4\over 8}~T^4
$$
\be
P_ {\rm tot}=p_4T^4+{1 \over 2}\dot\Phi^2={3r_4\over 8}~T^4.
\ee
Thus $\rho_{\rm tot}=3P_{\rm tot}$ which is the $4d$ equation of state
for thermal radiation.


\subsection{Numerical study}

\subsubsection{Without Wilson lines $y_5^a$}

Let us consider the case where the Wilson lines along the $X_5$-direction are switched off.
From Eq. (\ref{f}), we obtain \be p_4(u)={\Gamma({5\over2}) \over \pi^{5\over 2}} \left(
n_T^*f(u)+n_V^*\tf(u)\right),\quad \quad \tf(u)=u^3f(1/u)\, , \ee thanks to the
identities $n_T^*f(u)\equiv F(u,y_5^a=0)$ and $n_V^*\tf(u)\equiv \tF(u,y_5^a=0)$. Since $f+uf'$ and
$\tf+u\tf'$ vanish at the origin, Eq. (\ref{equ}) admits a universal solution $u=0$ for
arbitrary $n_V^*$ in the range $-n_T^*\le n_V^*\le n_T^*$. This solution corresponds to
$M(t)\equiv 0$ at some finite $T$, and is associated to a
$5$-dimensional purely thermal system ($R_5=\infty$).
The $4d$ effective description we have
considered is not valid in this case.

\noindent We are thus looking for non trivial solutions $u>0$ of Eq. (\ref{equ}),
which we write in the
form
\be 
\label{n} 
{n_V^*\over n_T^*}=-{f+uf'\over \tf+u\tf'}\, . 
\ee
One can show that such non-trivial solutions (and consequently, non-trivial
cosmological evolutions) exist only for models satisfying 
\be
 \label{cond1} -{1\over 15}<{n_V^*\over n_T^*}<0\, .
\ee 
The non vanishing root of Eq. (\ref{n}) is an increasing function of the ratio $n_V^*/ n_T^*$
satisfying $u\to +\infty$ when $n_V^*/n_T^*\overset{<}\to 0$ and $u\to 0$ when $n_V^*/n_T^*\overset{>}\to-{1/15}$, 
\begin{figure}[h!]
\begin{center}
\includegraphics[height=9cm]{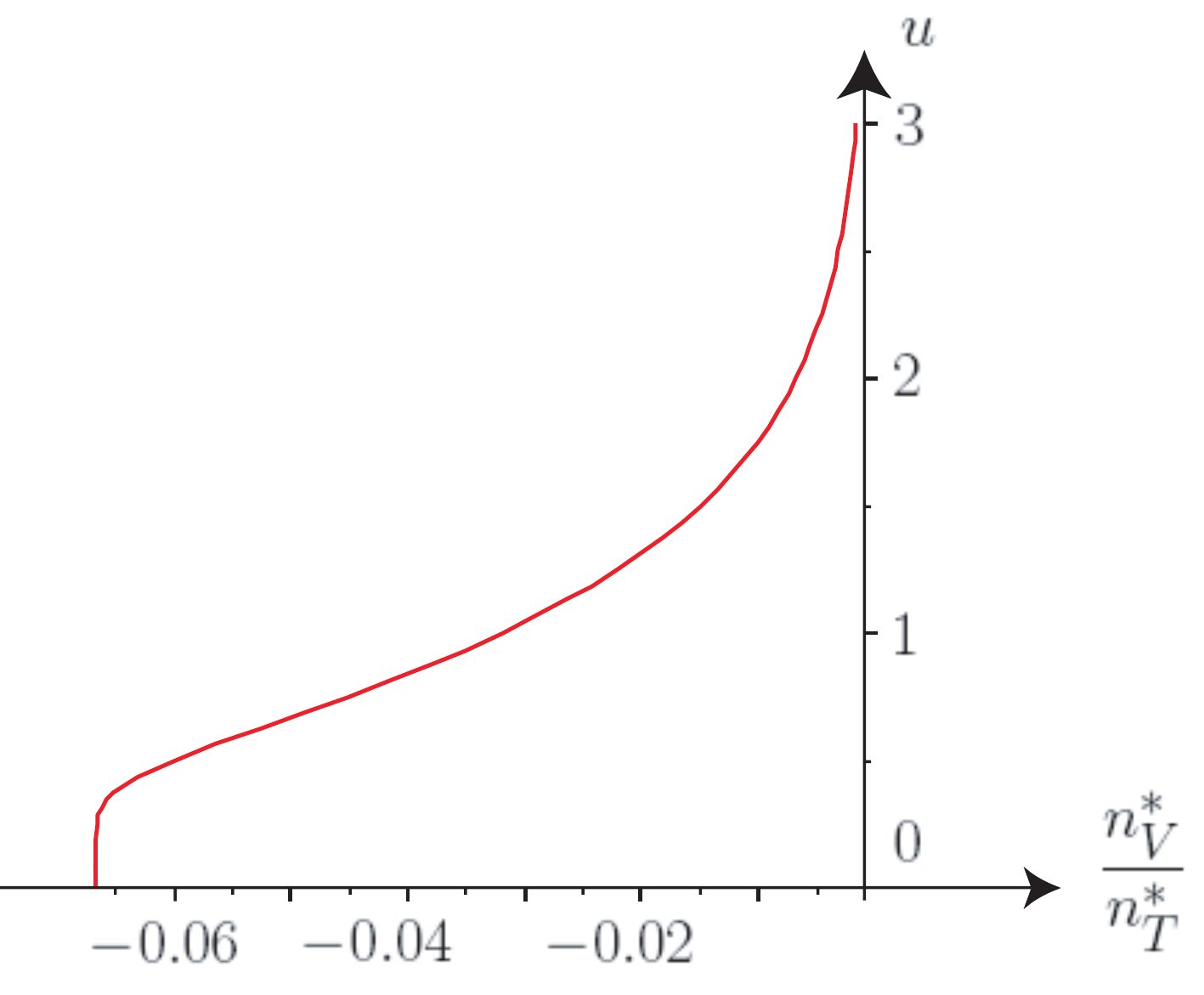}
\caption{\footnotesize \em The non trivial root $u$ of Eq. (\ref{equ}) as a function of the ratio $n_V^*/n_T^*$.}
\label{n_versus_u}
\end{center}
\end{figure} 
(See Fig. \ref{n_versus_u}).

\noindent The corresponding value of $c_r$ in Eq. (\ref{CReff}) is finite and always positive. This
can be seen by noting that the quantity $n_T^* f+n_V^*\tf$ appearing in the expression for $c_r$ equals
$-u(n_T^* f'+n_V^*\tf')>0$. The positivity follows since $f'$ and $\tf'$ are always negative and positive respectively.

\noindent When Wilson lines in the  $X_6$-direction are switched on, we can further study the sign of $\hk$. Defining $G(u,y_5^a=0)\equiv M_T^2 \,g(u) $ and $\tG(u,y_5^a=0)\equiv
M_V^{(2)}\,\tg(u)$, the quantity $p_2^R$ in (\ref{p2R}) becomes
\be
 p^R_2(u)=-{\Gamma({3\over2}) \over \pi^{3\over 2}}
\left( M_T^2(u)\,g(u)+M_V^{(2)}(u)\,\tg(u)\right),\quad \quad \tg(u)=ug(1/u)\, , 
\ee 
where $M_T^2(u)$, $M_V^{(2)}(u)$ are obtained from $M_T^2$ and $M_V^{(2)}$ under the replacement $(y_I^a/R_I)^2\to A_I^a c_I^a(u) Q_0^2$. Since $\hk$ and $p_2^R$ have same sign (see Eq. (\ref{keff})), one finds that
\be 
\label{Positk}
\hk\ge 0\quad \Longleftrightarrow \quad \left\{ u\ge 1~~\mbox{and}~~ -1\le {M_V^{(2)}(u)\over M_T^2(u)}\le -{g(u)\over \tilde g(u)}\right\}\; .
\ee
Note that a sufficient condition for $\hk$ to be negative is that $u\le 1$, or $M_V^{(2)}\ge 0$. For the particular value $u=1$, Eq. (\ref{n}) requires that $n_V^*/n_T^*\simeq -0.0320$.

\noindent Let us detail the models we considered in the previous
sections by computing the quantities $n_V^*$, $n_T^*$, $M_T^2$, $M_V^{(2)}$ to determine $u$ and $\hat k$. 

\noindent The $r^R_2+p^R_2=0$ constraint (\ref{r2p2})  ensures the  existence of non trivial Wilson lines and imposes extra restrictions on the space of critical solutions. Supposing for simplicity that the Wilson lines are functions of $u$ which are all proportional, $c^a_I(u)\equiv c(u)$, one finds on the critical solution with $u$ determined by $r_4=4p_4$ (Eq. (\ref{CReff}):
\be
\label{eta}
r_2^R+p_2^R=0\quad \Longleftrightarrow\quad 2-{d\log(g+s\tg)\over d\log u}=\eta \quad \where\quad s={M_V^{(2)}\over M_T^2}\; , \quad\eta={d\log c\over d\log u}\, ,
\ee
$s\in [-1,1]$ being a $u$-independent number in this case.
The above relation (\ref{eta}) reduces further  the permitted values of  $ s=M_V^{(2)}/M_T^2$. 
For $\eta$ small, one can show that  (\ref{eta}) is  satisfied  when $u$ is sufficiently large. The analysis for a generic value of $\eta$ is quite complex, and depends crucially on the particular choice of $y^a_I$ in the configuration space. In the explicit  examples we will consider below, we will skip this analysis. However, this is not a real obstruction since, as we will see later, the simultaneous presence of non trivial  Wilson lines in the directions 5 and 6 modifies the critical value of $u$ in a highly non trivial way. The study of the $\eta$-constraint on the  space of $(u,y_5^a)$  goes beyond the scope of this work. Keeping that  in mind we disregard the constraint when $y_5^a=0$ in the examples which are displayed below.

\noindent {\large \em Model 1:}\\
We consider an $N=4$ heterotic string model, for example the $E_8 \times E_8$
theory on $T^6$, at finite temperature and when supersymmetry is
spontaneously broken by choosing $Q_R=Q_F$. For this model
 \be n_T^* = n_V^*=2^3 \times 504\, . \ee Since $n_V^*>0$, there is no critical solution
$u>0$.

\noindent {\large \em Model 2:}\\
In the same $E_8 \times E_8$ heterotic string model,
the $R$-symmetry operator used to break supersymmetry is chosen to be
 $Q_F+Q_{E_8}+Q'_{E_8}$ in order to have
$n_V^*<0$ (see Eq. (\ref{nv-8}) for the $T=0$ case): \be n_T^* = 2^3 \times 504\,;\quad  n_V^* =
2^3\times (-8) \, .
 \ee 
 Then $n^*_V/n^*_T =-1/63\simeq -0.0159$, and so the model admits a critical
solution $u> 0$. One finds numerically that $u \simeq 1.46$ and $\gamma^4c_r\simeq 441$.

\noindent Let us consider Wilson lines $y_I^a$ in the direction $I= 6$ and define
 \be
 \label{WL}
(Y^a)^2={1\over 4\pi}  \left({y_6^a\over R_6}\right)^2,\;  a=1,...,16\, , 
\ee 
where
$a=1,...,8$ stand for the Cartan generators of the first $E_8$ factor and $a=9,...,16$ for the second.
The derivation of the charges $Q_a$ of the initially massless states (see the Appendix) gives:
\be M_T^2 =  2^3\times 60 \left( (Y^1)^2+\cdots +(Y^{16})^2\right)\, ,
\ee 
and 
\be 
M_V^{(2)} = 2^3\times (-4)\left( (Y^1)^2+\cdots+(Y^{16})^2\right)\, . 
\ee 
Their ratio
is Wilson line independent, $M_V^{(2)}/M_T^2=-1/15$, which is larger than $-g(u)/\tg(u)$, and so the effective curvature is negative. 

\noindent {\large \em Model 3:}\\
In the $N=2$ orbifold model (see section \ref{E7SU2} for details), we set
$Q_R=Q_F+Q_H+Q_{E_7}$ with $Q_{E_7}$ being 1 for the spinorial representations of $E_7$,
decomposed in terms of $SO(12)$ ones,
and 0 for the
vectorial ones. We find the following: 
\be n_T^* = 4064\,;\quad  n_V^* = 992 \, . 
\ee 
Since
$n_V^*/n_T^*$ is positive, this model doesn't admit a critical point $u> 0$  at this stage.

\noindent {\large \em Model 4:}\\
This is the $N=2$ orbifold model constructed in Sect. \ref{E7SU2}, with 
\be 
n_T^*=4064\;,\quad
n_V^*=-32\, . 
\ee 
Since $n_V^*/n_T^*=-1/127\simeq -7.87\cdot 10^{-3}$, there is again a non trivial
critical solution $u>0$. Numerically, we find $u\simeq 1.90$ and $\gamma^4c_r\simeq 130$.

\noindent Before switching on  Wilson lines in the direction $I=6$, the gauge group
is $E_7\times SU(2)\times E_8$. Let us consider arbitrary Wilson lines, $Y^{a}$, $a=1,...,16$:
$Y^1$,...,$Y^{6}$ for the Cartan generators of the $SO(12)$ subalgebra of $E_7$, $Y^7$ for the
Cartan generator of the  $SU(2)$ subalgebra of $E_7$, $Y^8$ for the $SU(2)$ factor and
$Y^9$,...,$Y^{16}$ for the $E_8$. The mass square scales, computed in the Appendix, are: 
\be 
M_T^2
=  624~\sum_{a=1}^6 (Y^a)^2+1248~(Y^7)^2+736~(Y^8)^2 +240~\sum_{a=9}^{16} (Y^a)^2 
\ee 
and 
\be
M_V^{(2)} =  -144~\sum_{a=1}^6 (Y^a)^2-288~(Y^7)^2+224~(Y^8)^2+240~\sum_{a=9}^{16} (Y^a)^2\, .
\ee
The smallest value of $M^{(2)}_V/M_T^2$ is reached when  any $Y^a$, $a=1,...,7$, is non
trivial (while $Y^a=0$, $a=8,...,16$). In this case $M^{(2)}_V/M_T^2=-3/13$, which is larger than $-g(u)/\tg(u)$, so that $\hk$ is negative. 


\subsubsection{With Wilson Lines $y_5^a$}

Any model originally characterized by the quantities $n_T^*$, $n_V^*$, $M_T^2$ and $M_V^{(2)}$ can
be deformed by switching on the Wilson lines $y_5^a$, $a=1,\dots,16$. We are looking for
solutions of Eq. (\ref{equ}) written in terms of the functions defined in (\ref{f}):
\be
F(u,y_5^a)+uF'(u,y_5^a)+\tF(u,y_5^a)+u\tF'(u,y_5^a)=0\, .
\ee 
We observe that the thermal contribution
described by $F(u,y_5^a)+uF'(u,y_5^a)$ vanishes at $u=0$ and is bounded for large $u$. In all of the
following examples, it is also positive. On the contrary, the
``effective potential'' corrections $\tF(u,y_5^a)+u\tF'(u,y_5^a)$ vanish at $u=0$ and diverge to $+\infty$ or
$-\infty$ at infinity. Thus, in presence of arbitrary Wilson
lines $y_5^a$, the universal solution $u=0$ remains and we are looking for non trivial ones $u>0$.
These can only arise when
$\tF(u,y_5^a)+u\tF'(u,y_5^a)$ takes negative values.

\noindent {\large \em Model 1:}\\
Among the Wilson lines $y_5^a$, ($a=1,...,8$ for the first $E_8$ factor and $a=9,...,16$ for the
second), we choose to switch on either \\
$i)$ $y_5^1$ (with $y_5^2=\cdots=y_5^{16}= 0$) or \\
$ii)$ $y_5^1=\cdots =y_5^8$ (with  $y_5^9=\cdots =y_5^{16}= 0$). \\
In these cases and at fixed Wilson lines, $\tF(u,y_5^a)+u\tF'(u,y_5^a)$ increases from $0$ to $+\infty$,
and so there is still no solution $u>0$.

\noindent {\large \em Model 2:}\\
For the Wilson lines defined in case $ii)$ above, one finds that the critical solution $u>0$
present before deformation is slightly shifted. However, in case $i)$, the root $u>0$ is sent to
$+\infty$ when $y_5^1$ approaches the numerical value $\simeq 0.227$ from below. For $y_5^1$ above this
critical bound, there is no non trivial solution anymore.

\noindent {\large \em Model 3:}\\
As said before, this model does not admit a critical solution $u>0$ when all Wilson lines in the
fifth dimension are switched off. We consider four patterns of deformations, by switching on a Wilson
line for a single Cartan generator:\\
$i)$ $y_5^1$ (Cartan generator of $SO(12)\subset E_7$)\\
$ii)$ $y_5^7$ (Cartan generator of $SU(2)\subset E_7$)\\
$iii)$ $y_5^8$ (Cartan generator of $SU(2)$)\\
$iv)$ $y_5^9$ (Cartan generator of $E_8$)

\noindent In cases $i)$ and $iv)$, a large enough Wilson line deformation generates a non-trivial
solution. We find the critical bound $y^1_5\simeq 0.895$ for case $i)$, and $y^9_5\simeq 0.772$ for case $iv)$. In
all these cases, the phase transition occurs when the limit at $u\to \infty$ of the
``effective potential'' contribution $\tF(u,y_5^a)+u\tF'(u,y_5^a)$ is switched from $+\infty$ to $-\infty$.
In some sense, a solution $u=+\infty$ appears at the transition, and then decreases for a larger Wilson
line, and finally reaches a non-zero minimal value. In cases $ii)$ and $iii)$, since
$\tF(u,y_5^a)+u\tF'(u,y_5^a)$ is positive for any value of the Wilson line, the deformation does not create a non-trivial solution.

\noindent {\large \em Model 4:}\\
We consider the same four patterns of Wilson lines. This time, the critical solution $u>0$ exists before
deforming the model. We find that in the above cases, switching on a large enough  Wilson line makes this
solution disappear. However, two distinct behaviors are found:\\
In cases $i)$ and $ii)$, we find again that the phase transition corresponds to a change in the
behavior of the ``effective potential'' contribution at $\infty$. In a similar way as before, the
solution $u>0$ existing before deformation will increase when switching on the Wilson line, then go to
$+\infty$ at the transition point, and disappear for a larger Wilson line. We find the critical
bound $y_5^1\simeq 0.105$ for case $i)$, and $y_5^7\simeq 0.074$ for case $ii)$.\\
For  $iii)$ and $iv)$, switching on the Wilson line makes the critical solution $u>0$ decrease towards zero, so that we are left only with the universal solution when the Wilson line is above some bound. In case $iii)$, this maximal value is $y_5^8\simeq 0.205$, while it is $y_5^9\simeq 0.207$ for case $iv)$.

\noindent Some remarks are in order\\
\indent $\bullet$ In all cases presented above, when a critical solution $u>0$ exists, the effective radiation term $c_r$ given
by Eq. (\ref{CReff}) is strictly positive.\\
\indent $\bullet$ 
In some models, incorporating the Wilson lines in the fifth direction allows the critical solution $u>0$ to be close to 0
or large. In other words, a hierarchy between
the scales $M$ and $T$ can be found by tuning the moduli $y_5^a$.


\section{Cosmological evolutions}

When a non degenerate solution $u > 0$ exists, the model admits a well defined low energy description in four dimensions.
The dynamics in this regime is controlled by the Friedmann-Hubble equation,
whose behavior depends drastically on the signs of $c_r$ and $\hk$.
In all models we considered here, with initial $N=4$ or $N=2$ supersymmetry,
$c_r$ turns out to be positive.
However, this situation may not be generic in more complex stringy examples with initial $N= 1$ supersymmetry.
For completeness, we briefly describe all possible cosmologies arising for any positive or negative value of $c_r$.
In addition, we allow non trivial time dependent profiles for the moduli $\phi_i$, giving $c_m> 0$.

\subsection{Solutions for $c_r>0$}

\noindent $\bullet$ {\em \large For $\hk=0$, $c_r>0$, $c_m\ge0$}

\begin{figure}[h!]
\begin{center}
\includegraphics[height=7cm]{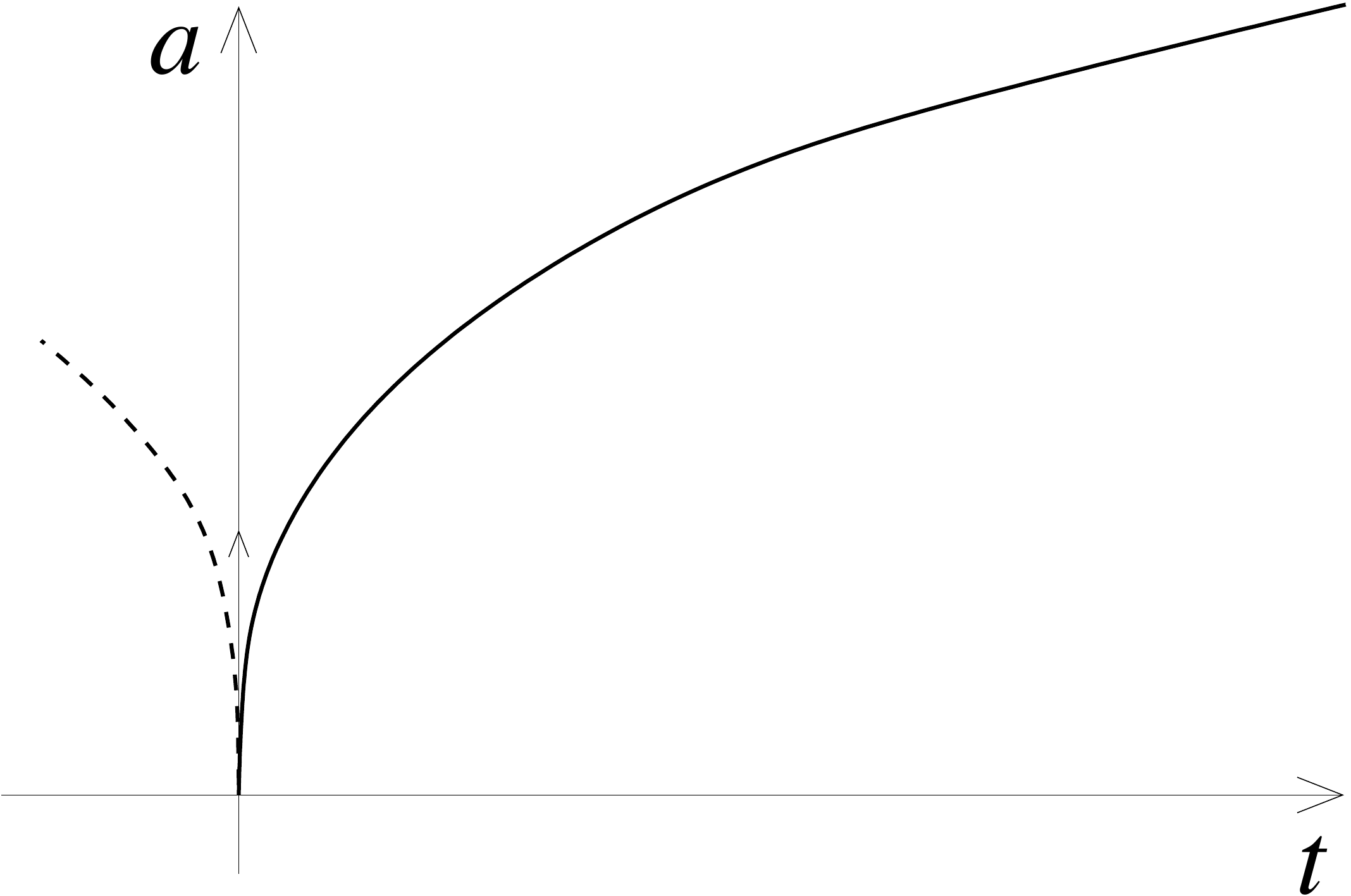}
\caption{\footnotesize \em Cosmological evolution for the case $\hk=0$, $c_r>0$, $c_m\ge 0$.}
\label{fig_k0_rp_mp}
\end{center}
\end{figure}

\noindent When Wilson lines in the directions  $6, 7,\dots, 10$ are absent, the curvature $\hk$ vanishes.
In real time, the Friedmann-Hubble Eq. (\ref{cosmoeff}),
\be
\label{Friedman1}
3 H^2= {c_r \over a^4}+{c_m \over a^6}\, ,
\ee
can be used to express the time variable $t$ as an integral function of the
scale factor $a$,
\be
\label{t1}
t(a)=\sqrt{{3\over c_r}} \int_0^a {v^2 dv\over \sqrt{v^2+a_0^2}}\, , \quad  (a\ge 0),\quad \quad a_0=\sqrt{{c_m\over c_r}}\, .
\ee
In this form, it is straightforward to draw $a(t)$ (see Fig. \ref{fig_k0_rp_mp}). The explicit solution is
\be
t(a)= t_0\left({a\over a_0} \sqrt{1+{a^2\over a_0^2}} - \mbox{arcsinh}\left({a\over
a_0}\right)\right),\quad \quad t_0={\sqrt{3}\over2}\, {c_m\over c_r^{3/2}}\, ,
\ee
describing an expanding universe that starts with a big bang. As can be seen from Eq. 
(\ref{Friedman1}), the slope $\dot a$ is infinite when $a$ vanishes. At large $t$, the scale factor
behaves as in the $c_m=0$ particular case:
\be
a(t)= \left({c_r\over 3}\right)^{1/4}\sqrt{2t}\; , \quad  (t\ge 0)\, .
\ee

\noindent Since the transformation $t\to -t$ is a symmetry of the Friedmann-Hubble equation, the previous expanding
solutions have contracting counterparts and thus ending at $t=0$ with a
big crunch. Finally, since the RHS  of Eq. (\ref{Friedman1}) is positive, there is no solution in
Euclidean time.

\noindent The effective field theory description always breaks down before the occurrence of a
space-like singularity, when
the temperature (in string frame) is of order the Hagedorn temperature. At this temperature scale, new
stringy dynamics must be taken into consideration which can result into a phase transition,
realizing the scenario of \cite{vafabrand}.

\noindent $\bullet$ {\em \large For $\hk>0$, $c_r>0$, $c_m\ge0$}

\noindent When we switch on Wilson lines we could have $\hk>0$. Rewriting the Friedmann-Hubble Eq. (\ref{cosmoeff})
 in the form \be \label{H5} (a^2\ad)^2=\hk
(a^2+a_-^2)(a_+^2-a^2),\quad \quad a_\pm=\sqrt{\sqrt{{c_r^2+12\hk c_m}}\pm c_r \over
6\hk}\, ,
\ee
one expects a cosmological evolution satisfying $0\le a\le a_+$ should exist, while a
solution with a scale factor greater than $a_+$ should make sense in imaginary time only. In real
time, one can actually express $t$ as a function of $a$ as follows
\be
\label{t5}
t(a)=\pm \left(t_i+{1\over \sqrt{\kh}}\int_0^a {v^2dv\over\sqrt{ (v^2+a_-^2)(a_+^2-v^2)}}\right)\, ,\quad (0\le a\le a_+)\, , \ee where \be t_i=-{1\over \sqrt{\kh}}\int_0^{a_+} {v^2dv\over\sqrt{
(v^2+a_-^2)(a_+^2-v^2)}}\, .
\ee
From these expressions, one can see that the cosmological
evolution  starts with a big bang at $t=t_i$. It expands until $t=0$ where the maximum size of
the universe $a_+$ is reached, and then it contracts until a big crunch occurs at $t=-t_i$.

\begin{figure}[h!]
\begin{center}
\includegraphics[height=8cm]{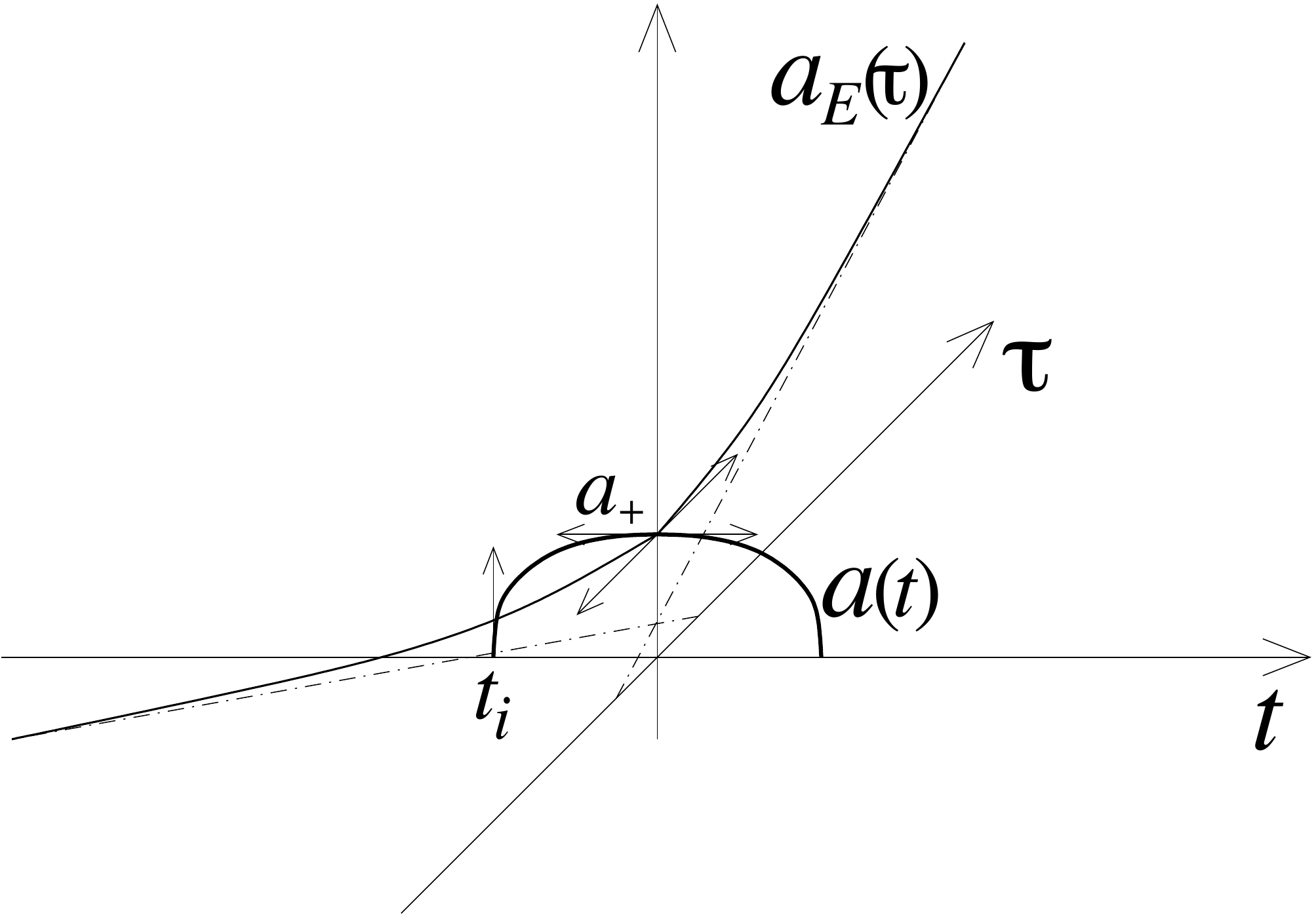}
\caption{\footnotesize \em Cosmological evolution for the case $\hk>0$, $c_r>0$, $c_m\ge 0$ (in bold line). A big bang and a big crunch are occurring at $t=t_i$ and $t=-t_i$ respectively. This solution is connected to a Euclidean one at $t=-i\t=0$ that is asymptotically linear.}
\label{fig_kp_rp_mp}
\end{center}
\end{figure}

\noindent To find a Euclidean solution, one needs to
consider a scale factor greater than $a_+$. It is then possible to find it by proceeding as before,
or use the fact that such a solution can be obtained by analytic continuation of the expression
(\ref{t5}) at $t=0$, where $a=a_+$. One can write
\be
t=\pm \left( t_i+{1\over \sqrt{\kh}}\int_0^{a_+} {v^2dv\over\sqrt{(v^2+a_-^2)(a_+^2-v^2)}}+
{1\over \sqrt{\kh}}\int_{a_+}^{a_E} {v^2dv\over\sqrt{-(v^2+a_-^2)(v^2-a_+^2)}}\right)\equiv -i\t \, , \ee
from which we derive
\be
\label{t6}
\t(a_E)=\pm {1\over \sqrt{\kh}}\int_{a_+}^{a_E}
{v^2dv\over\sqrt{(v^2+a_-^2)(v^2-a_+^2)}}\, , \quad  (a_E\ge a_+)\, .
\ee
Fig. \ref{fig_kp_rp_mp} represents the solutions $a(t)$ and $a_E(\t)$.

\noindent  In the particular case where $c_m=0$,
the solutions (\ref{t5}) and (\ref{t6}) are taking the explicit forms
\be
a(t)=a_+\sqrt{1-\left({t\over t_i}\right)^2} \; ,   \quad (t_i\le t\le -t_i),   \quad
\quad a_+=\sqrt{{c_r\over 3\hk}}\; , \quad  t_i=-{1\over \hk}\sqrt{{c_r\over 3}}\, ,
\ee
and
\be
a_E(\t)=a_+\sqrt{1+\left({\t\over t_i}\right)^2} \, ,
\ee
whose shapes are similar to the generic case with $c_m>0$.

\noindent $\bullet$ {\em \large For $\hk<0$, $c_r> 0$, $c_m\ge0$}

\begin{figure}[h!]
\begin{center}
\includegraphics[height=7cm]{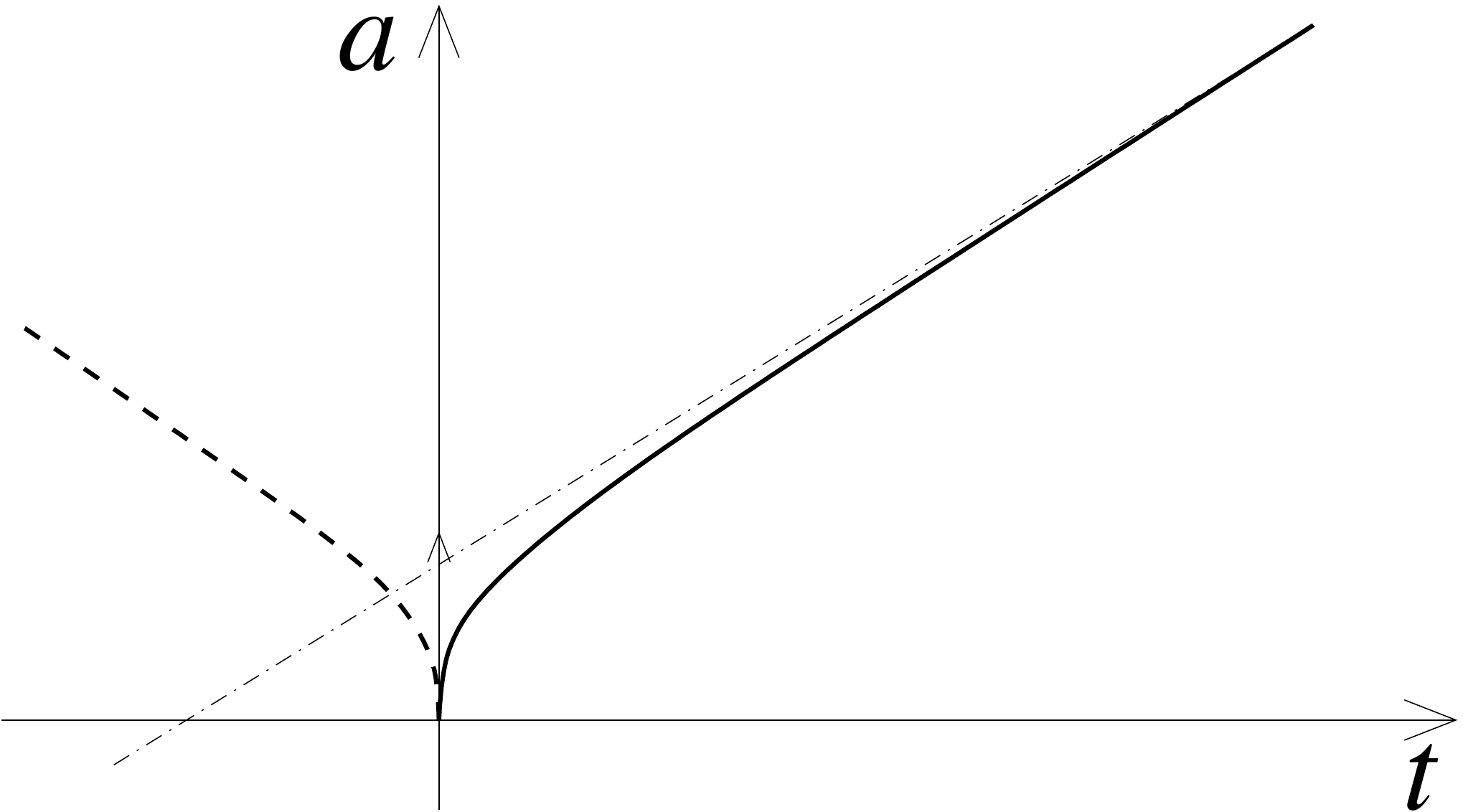}
\caption{\footnotesize \em Cosmological evolution for the case $\hk<0$, $c_r>0$, $c_m\ge 0$.}
\label{fig_kn_rp_mp}
\end{center}
\end{figure}

\noindent   This case is easier to deal with. Eq. (\ref{cosmoeff}) can be rewritten as
\be
\label{H6}
(a^2\ad)^2=|\hk| (a^2+a_-^2)(a^2+a_+^2)\; , \quad \quad a_\pm=\sqrt{{c_r\pm \sqrt{c_r^2-12|\hk| c_m} \over 6| \hk| }}\, ,
\ee
and admits the expanding solution
\be
\label{t7}
t(a)={1\over \sqrt{\kh}}\int_{0}^{a} {v^2dv\over\sqrt{(v^2+a_-^2)(v^2+a_+^2)}}\,.
\ee
(See Fig. \ref{fig_kn_rp_mp}.) After a big bang,  the scale factor is growing linearly in time.

\noindent The result for the particular case $c_m=0$ can be written more explicitly. The solution takes the form
\be
a(t)=a_+\sqrt{\left({t+t_0\over t_0}\right)^2-1}\; ,
 \quad  (t\ge 0),\quad \quad a_+=\sqrt{{c_r\over 3|\hk|}}\, , \quad t_0={1\over |\hk|}\sqrt{{c_r\over 3}}\, .
\ee
These cosmological solutions do not admit a sensible Euclidean continuation.

\subsection{Exotic cosmologies with $c_r<0$}

Although in all explicit models we presented before $c_r$ is positive,
it is interesting to analyze the exotic situation with negative $c_r$,
which is not a priori forbidden in more general cases with $N=1$ initial supersymmetry.

\noindent $\bullet$ {\em \large For $\hk=0$, $c_r< 0$, $c_m\ge0$}

\begin{figure}[h!]
\begin{center}
\includegraphics[height=8cm]{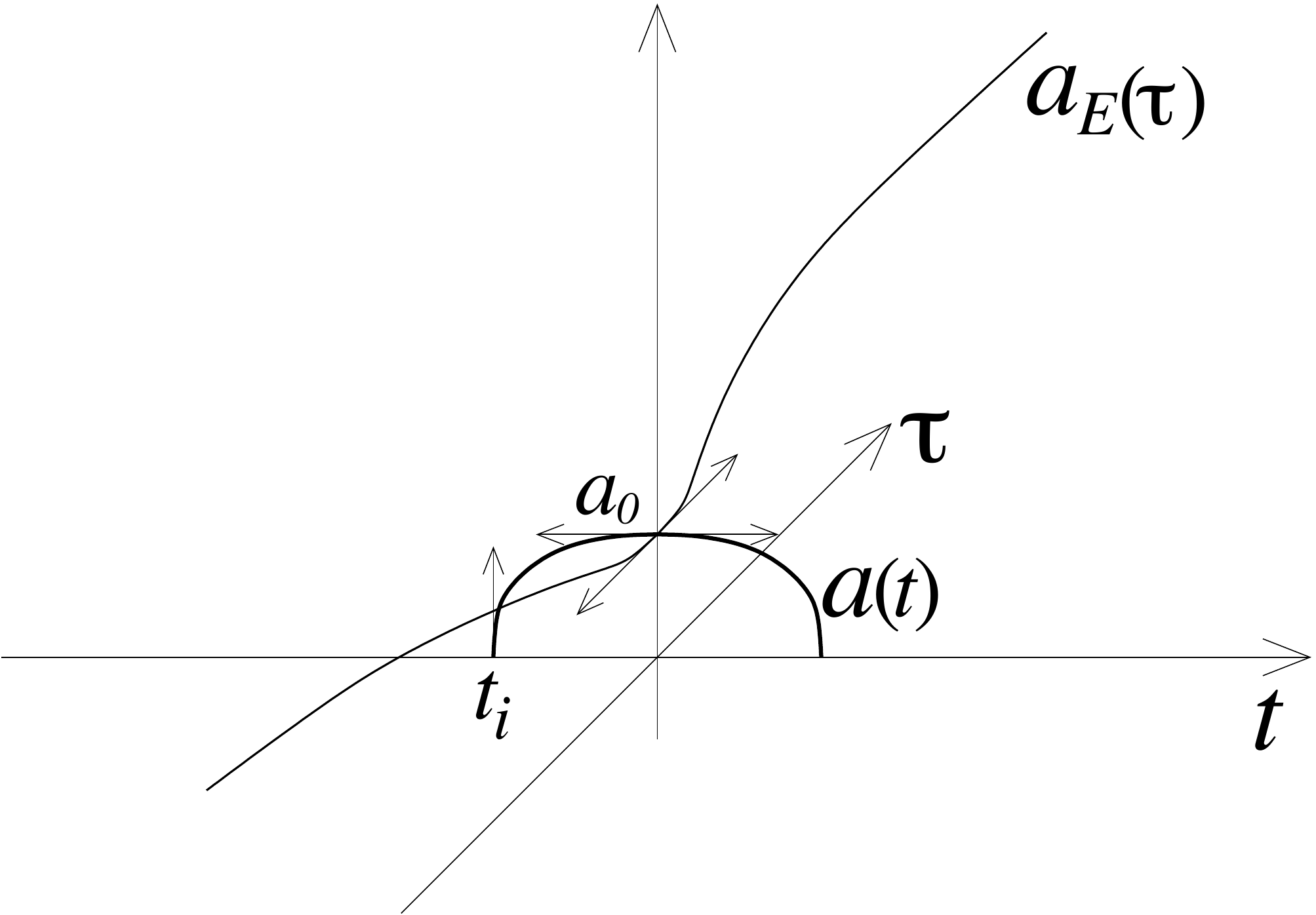}
\caption{\footnotesize \em Cosmological evolution for the case $\hk=0$, $c_r<0$, $c_m\ge 0$ (in bold line). A big bang and a big crunch are occurring at $t=t_i$ and $t=-t_i$ respectively. This solution is connected to a Euclidean one at $t=-i\t=0$.}
\label{fig_k0_rn_mp}
\end{center}
\end{figure}

\noindent A cosmological evolution in real time only exists if $c_m$ is switched on.
The scale factor satisfies $0\le a\le a_0$ where
\be
a_0=\sqrt{{c_m\over |c_r|}}\, .
\ee
Between a big bang at $t_i<0$ and a big crunch at $-t_i>0$, $a$ reaches a maximum $a_0$ at $t=0$.
At this time, an analytic continuation is allowed: A Euclidean solution satisfies $a_E\ge a_0$ and goes
to infinity for large positive or negative Euclidean time (see Fig. \ref{fig_k0_rn_mp}).

\noindent $\bullet$ {\em \large For $\hk>0$, $c_r\le0$, $c_m\ge0$}

\begin{figure}[h!]
\begin{center}
\includegraphics[height=8cm]{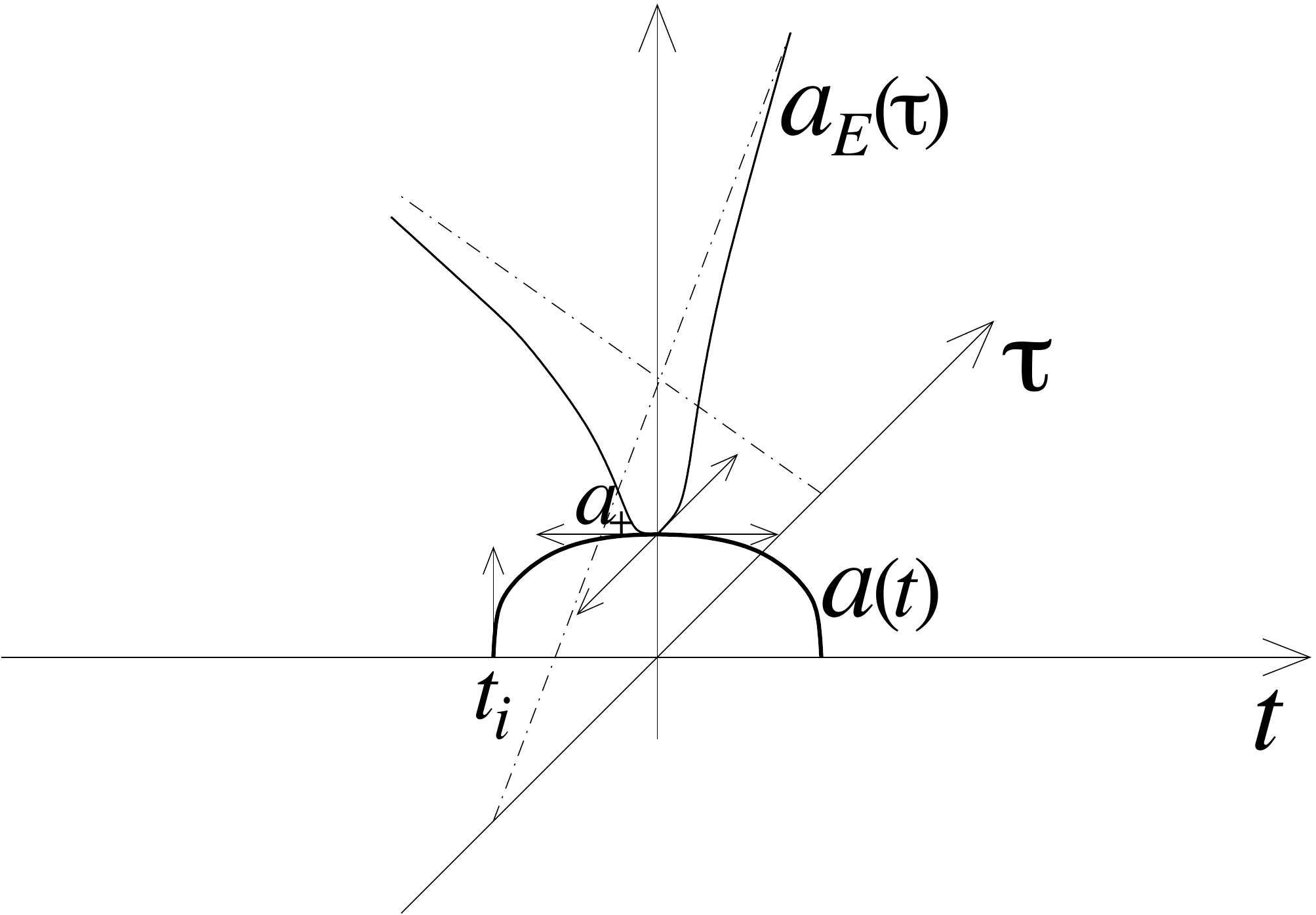}
\caption{\footnotesize \em Cosmological evolution for the case $\hk>0$, $c_r\le 0$, $c_m> 0$ (in bold line). A big bang and a big crunch are occurring at $t=t_i$ and $t=-t_i$ respectively. This solution is connected to a Euclidean one at $t=-i\t=0$ that is asymptotically linear and has two symmetric inflexion points.}
\label{fig_kp_rn_mp}
\end{center}
\end{figure}

\noindent Most of the considerations of this case are identical to the one derived for $\hk>0$, $c_r>0$, $c_m\ge0$. In particular, the Friedmann-Hubble equation is still given by Eq. (\ref{H5})
and both the cosmological solution (\ref{t5}) and the Euclidean one (\ref{t6}) are valid.
They are shown in Fig. \ref{fig_kp_rn_mp}.
As long as $c_m>0$, the only qualitative difference with the case $c_r>0$
is that the Euclidean solution has two symmetric inflexion points.
However, when $c_m$ vanishes, the evolution in real time ceases to exist.

\noindent $\bullet$ {\em \large For $\hk<0$, $c_r< 0$, $c_m\ge0$}

\noindent   This case presents the most interesting features and involves either a first or second order phase transition in the early universe. The former case is the only one considered in this paper, where a Euclidean solution has a finite action and thus can be interpreted as an instanton involved in a tunneling effect. These behaviors are qualitatively similar to the inflationary case studied in \cite{KPThermal, KP}. To be more specific we consider the Friedmann-Hubble  equation in the form
\be
\label{H7}
3(a^2\dot a)=3|\hk|a^4-|c_r|a^2+c_m\, ,
\ee
and discuss various regimes, depending on the value of the discriminant of the RHS
\be
\label{discri}
\delta\equiv c_r^2-12|\hk| c_m\, .
\ee

\begin{figure}[h!]
\begin{center}
\includegraphics[height=9cm]{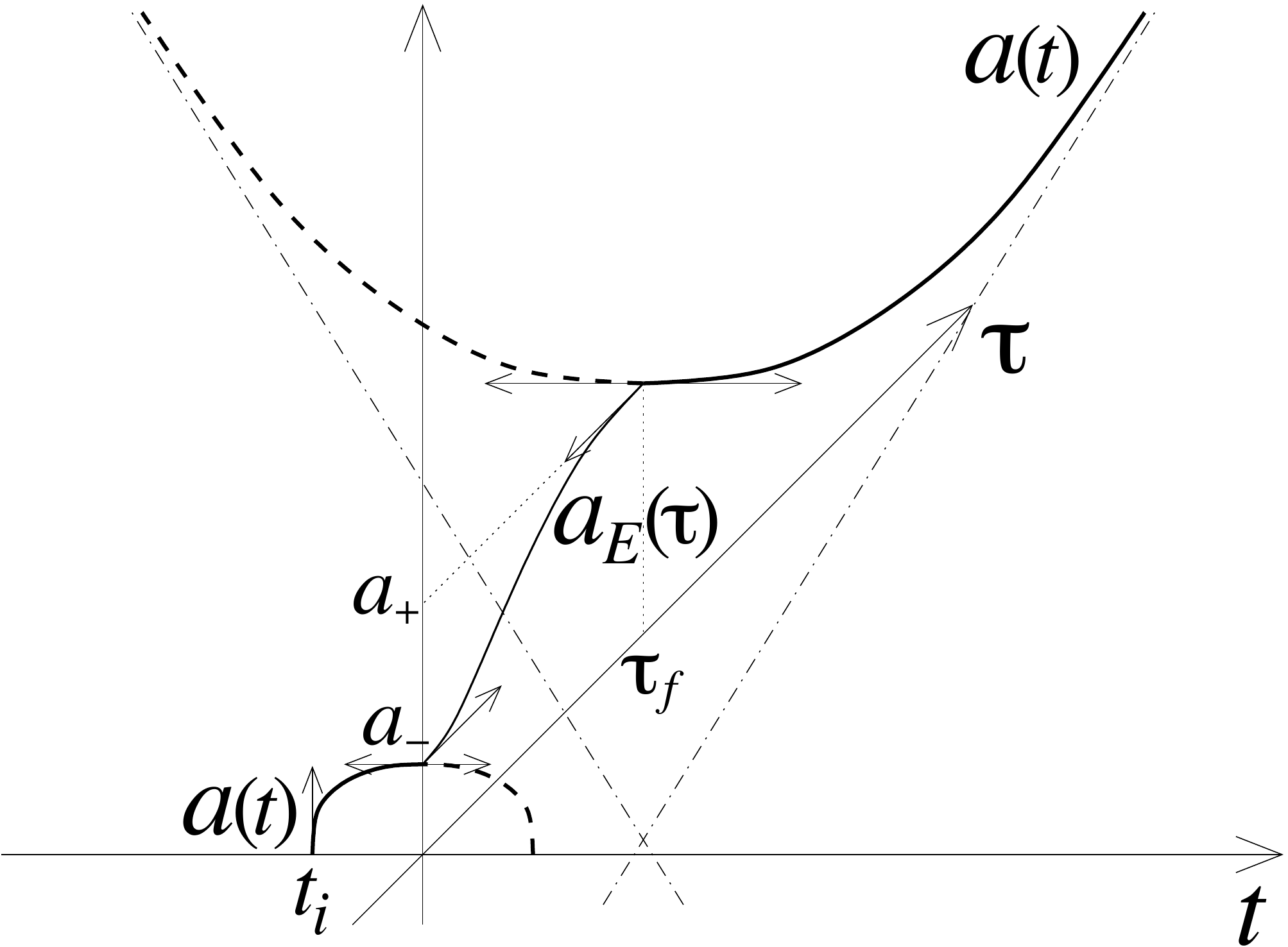}
\caption{\footnotesize \em There are two cosmological evolutions (in bold lines) for the case $\hk<0$, $c_r\le0$, $c_m\ge 0$, when $\delta>0$. The first one starts with a big bang at $t=t_i$ and ends with a big crunch at $t=-t_i$. The second one has a contracting phase followed by an expanding one. These two branches are connected to each other by a first order phase transition via an instanton.}
\label{fig_kn_rn_mp_dp}
\end{center}
\end{figure}

\noindent $i)$ When $\d>0$, there are two critical values for the scale factor
\be
a_\pm=\sqrt{{|c_r|\pm \sqrt{c_r^2-12|\hk| c_m} \over 6| \hk| }}\, .
\ee
Eq. (\ref{H7}) then admits two distinct cosmological evolutions. The first one satisfies $0\le a\le a_-$ and corresponds to the usual dynamics between a big bang at $t_i<0$ and a big crunch at $-t_i$. The scale factor reaches a maximum $a_-$ at $t=0$. The second one describes an asymptotically linear contracting solution followed by an asymptotically linear expanding one. The two branches are smoothly connected at $t=0$, where the scale factor reaches a minimum value $a_+$.
Therefore, this solution is non-singular. (See Fig. \ref{fig_kn_rn_mp_dp}.)

\noindent The two cosmological evolutions are also related to one another by a double analytic continuation: $t=-i\t$ and then $\t=\t_f+it$. Between $\t=0$ and $\t=\t_f$, a Euclidean solution whose action can be shown to be finite is allowed. It is thus an instanton between the two branches in real time and contributes to a first order phase transition. We note that when $c_m=0$, the big bang / big crunch solution disappears since $a_-$ vanishes. 

\begin{figure}[h!]
\begin{center}
\includegraphics[height=8cm]{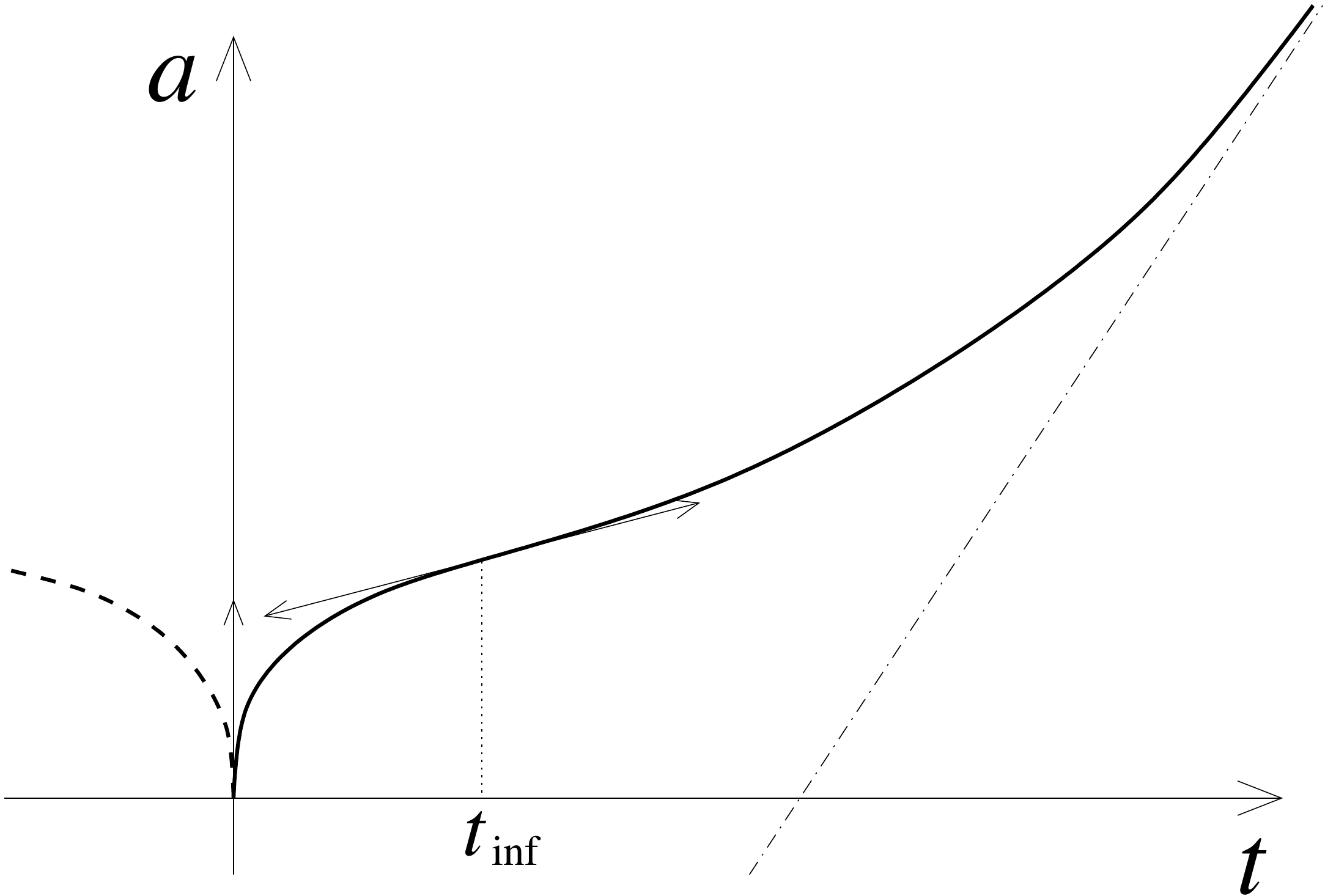}
\caption{\footnotesize \em Cosmological evolution for the case $\hk<0$, $c_r\le0$, $c_m\ge 0$, when $\delta<0$. It describes a second order phase transition occurring at $t_{\mbox{\tiny inf}}$, between a phase that starts with a big bang to another phase that expands linearly in time.}
\label{fig_kn_rn_mp_dn}
\end{center}
\end{figure}

\noindent $ii)$ Let us turn now to the second case where the discriminant (\ref{discri}) satisfies $\d<0$. Eq. (\ref{H7}) does not admit any critical point and the scale factor is never stationary.  There is a single cosmological evolution (and no Euclidean solution). It increases from a big bang at $t=0$, while for large $t$, its time dependence becomes linear.
Thus, close to the big bang, the evolution is similar to the first solution occurring when $\d>0$, while for large $t$ its behavior is similar to the second expanding solution. Since there is an inflexion point at $t=t_{\mbox{\tiny inf}}$ when $a=\sqrt{2c_m/|c_r|}$, the cosmological evolution for $\d<0$ can be interpreted as a second order phase transition between the same initial and final states encountered in the first order phase transition for $\delta> 0$. (See Fig. \ref{fig_kn_rn_mp_dn}.)

\begin{figure}[h!]
\begin{center}
\includegraphics[height=8cm]{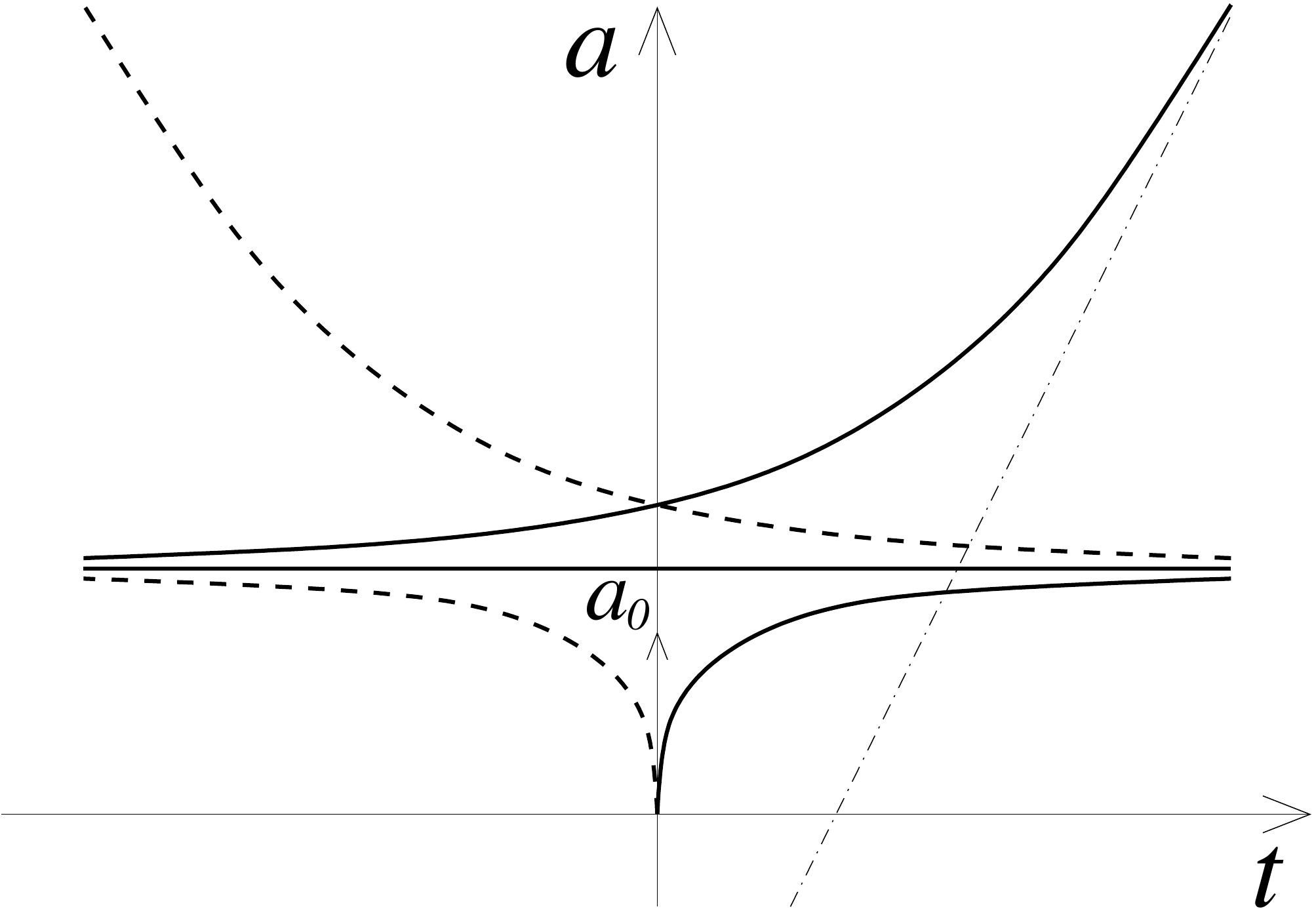}
\caption{\footnotesize \em There are two expanding (contracting) cosmological evolutions for the case $\hk<0$, $c_r\le0$, $c_m\ge 0$, when $\delta=0$. All are asymptotic to the static solution $a\equiv a_0$.}
\label{fig_kn_rn_mp_d0}
\end{center}
\end{figure}

\noindent $iii)$ In the critical case $\d=0$, Eq. (\ref{H7}) admits two expanding cosmological evolutions which are asymptotic to a static one, $a\equiv a_0$, where
\be
a_0\equiv a_\pm=\sqrt{{|c_r|\over 6|\hk|}}=a_{\mbox{\tiny inf}}=\sqrt{{2c_m\over |c_r|}}\, ,
\ee
together with two contracting ones obtained by time reversal. The first expanding solution starts with a big bang, while the second one is linear for large positive time. (See Fig. \ref{fig_kn_rn_mp_d0}.)
%


\section{Conclusions}
We have obtained several cosmological solutions in a large class of
four dimensional heterotic
string compactifications with spontaneously broken $N=4$ or $N=2$ space-time supersymmetry.
The cosmological
evolution is induced once radiative quantum and thermal corrections are taken into
consideration. These corrections are calculated at the perturbative string level and shown
to possess universal scaling properties. The reason is an underlying duality between
the temperature and the supersymmetry breaking scale.

\noindent Our solutions correspond to homogeneous and isotropic Friedmann-Robertson-Walker
universes.
They are characterized by the ratio of the supersymmetry breaking
scale to the temperature, and this ratio remains constant during time evolution.
Even though Kaluza-Klein states associated to the supersymmetry breaking cycle are thermally
excited, the equation of state governing cosmological evolution is identical to that
of massless thermal radiation in four dimensions. This is due to the special relation between the no-scale modulus field
associated to the supersymmetry breaking scale and the Hubble parameter:
$\dot \Phi^2 = 2H^2/3$.
Universes with spherical, toroidal or hyperbolic spatial sections can be found once we incorporate Wilson
line deformations.

\noindent In this paper we focused on the low temperature phase of the models. When the temperature is close to
the Hagedorn temperature our effective field theory analysis breaks down and new stringy dynamics must
be taken into consideration. It would be interesting to investigate if phase transitions can occur in
these models as the temperature approaches the Hagedorn temperature, and whether
such phase transitions result in non-singular time-dependent geometries. To this extent
it could prove useful to incorporate in our work the proposal of \cite{akd}, where such a
phase transition is shown to occur in $N=4$ heterotic string models, and study
the cosmological implications.

\noindent It would be interesting to extend our analysis to the $N=1$ heterotic orbifold models,
and for the cases where supersymmetry is broken spontaneously. In this class of models, one expects
to find inflationary phases, once radiative and thermal corrections are properly
taken into account. The analysis of \cite{KP,KPThermal} reveals interesting transitions between such
inflationary phases and radiation dominated phases with similar properties to those found in this work.
In our examples, the coefficient of the $1/a^4$ term in the Friedmann-Hubble   equation is positive.
Perhaps among the $N=1$ examples it is possible to find models characterized by negative values
of this coefficient. Then non trivial cosmological phenomena would occur,
including first or second order phase transitions that allow for the possibility to realize the
proposal for the creation of a universe from ``nothing'' \cite{WFU} in string theory \cite{KP, KPThermal}.

\noindent The relation between the supersymmetry breaking scale with the temperature is a key property
of our solutions. Suppose that such a scaling property persisted in an early universe epoch, and that
initially supersymmetry was broken around the string scale. During such epoch, the
Susy-breaking scale gets lower and lower as the universe expands and cools. At lower temperatures new
dynamics may become relevant that can stabilize this scale. Such a scenario can give us a new perspective
on how to handle the hierarchy and naturalness problems.

\vspace{.5cm}
\noindent {\em Note added:} A follow up of the present work can be found in \cite{BKP}. The radiation era we have described is not only obtained for specific initial boundary conditions that select the critical trajectory. Instead, the radiation era is an attractor of the dynamics. It is reached asymptotically  for generic initial boundary conditions.


\section*{Acknowledgements}

We are grateful to Constantin Bachas, Ramy Brustein, Dieter L\"ust,  Marios Petropoulos, Jan Troost and Fabio Zwirner
for useful discussions.
N.T. thanks the Ecole Normale Superieure and C.K. and H.P. the University of Cyprus for hospitality.\\
\noindent The work of C.K. and H.P. is partially supported by the EU contract MRTN-CT-2004-005104
and the ANR (CNRS-USAR) contract  05-BLAN-0079-01 (01/12/05). N.T. and C.K. are supported by the EU
contract MRTN-CT-2004-512194. H.P. is also supported by the EU contracts MRTN-CT-2004-503369 and
MEXT-CT-2003-509661, INTAS grant 03-51-6346, and CNRS PICS 2530, 3059 and 3747, while N.T. is also supported
by an INTERREG IIIA Crete/Cyprus program.

\newpage

\begin{center}
{\Large\bf Appendix A}
\end{center}
\renewcommand{\theequation}{A.\arabic{equation}}
\renewcommand{\thesection}{A.}
\setcounter{equation}{0}

\noindent In this Appendix, we provide details on the derivation of the charges associated to the Cartan generators of the gauge groups of the  heterotic string models we considered. We also derive the contributions to $M_T^2$ and $M_V^{(2)}$ (see Eq. (\ref{P})) associated to  Wilson lines in the internal direction 6 only. Since the result is linear in the  $(y^i_6)^2$'s, the  result for arbitrary Wilson lines in the directions 6, 7, 8, 9, 10 is obtained under the replacement   $(y^i_6)^2/(4\pi R^2_6)\rightarrow\sum_I (y^i_I)^2/(4\pi R^2_I)$.

\noindent Consider the $N=4$ heterotic model with gauge group $E_8\times E_8$.
When built out from $32$ worldsheet fermions as in the standard procedure,
the vertex operators for the gauge fields associated to the Cartan generators of the first $E_8$
factor are given by \be \lambda^{2i-1}\lambda^{2i},~~~i=1,\dots, 8,\ee whereas those associated
to the Cartan generators of the second $E_8$ are given by
\be \lambda^{2i-1}\lambda^{2i},~~~i=9,\dots, 16.\ee
In terms of representations of $SO(16)$,
the adjoint representation of $E_8$
decomposes as \be{\bf 248}~=~ {\bf 120}\oplus{\bf 128},\ee where the {\bf 120} is the adjoint representation
of $SO(16)$ and the {\bf 128} is the spinorial representation with positive chirality. The
corresponding vertex operators can be written explicitly by bosonizing the $32$ fermions into 16 free bosons
$H_i$, $i=1,\dots, 16$. This formalism has the advantage to make the roots of the Lie
algebra appear in a clear way. For the {\bf 120} we obtain
\be {\bf 120}~:~\left(e^{i(\pm H_j\pm H_k)},~~(j\neq k)\right)  ~\oplus~i\partial H_j,~~~j,k \in
\{1,\dots,8\}.\ee The $8$ latter vertex operators correspond to the Cartan generators. For the {\bf 128} we have \be {\bf
128}~:~~e^{\frac i2(\epsilon_1 H_1+\epsilon_2 H_2+\cdots+\epsilon_8 H_8)},\ee with the GSO constraint
$\prod_{i=1}^8 \epsilon_i=1$. Our goal here is to compute the quantity 
\be 
M_T^2 ={1\over 4\pi R_6^2}
\sum_{s\in{\bf 248}} (Q_a^{s}y^a_6)^2,\ee where we denote by $Q_a^{s}$ the charges of a state $s$ with
respect to the Cartan generators.
We are going to consider the case of one $E_8$ gauge group; the
generalization to $E_8\times E_8$
is straightforward.

\noindent To begin with, we recall that the Cartan states are  neutral. Then we are interested
in the 112 remaining states of the {\bf 120} adjoint representation.
It is not hard to see that \be \sum_{s\in {\bf 120}}
(Q_a^{s} y^a_6)^2 = \sum_{i\neq j \in \{1,\dots, 8\}}~\sum_{\epsilon_1,\epsilon_2 = \pm 1} (\epsilon_1
y^i_6+\epsilon_2 y^j_6)^2.\ee Therefore we get a quadratic polynomial in the $y^i_6$'s. Noticing
that
this polynomial is invariant under the transformations $y^i_6 \leftrightarrow -y^i_6$ and $y^i_6
\leftrightarrow y^j_6$, we obtain\be \sum_{s\in {\bf 120}} (Q_a^{s} y^a_6)^2 = \alpha\sum_{i=1}^8
(y^i_6)^2.\ee Computing  the $(y^1_6)^2$ term, we get $\alpha=28$.

\noindent For the {\bf 128}, we see
that \be \sum_{s\in{\bf 128}} (Q_a^{s}y^a_6)^2 = \sum_{\epsilon_1,\dots,\epsilon_7=\pm 1}\frac
14\left(\epsilon_1 y^1_6 + \epsilon_2y^2_6 +\cdots + (\prod_{i=1}^7\epsilon_i)
y^8_6\right)^2.\ee If we
set $y^8_6=0$, the symmetries $y^i_6 \leftrightarrow -y^i_6$ and $y^i_6 \leftrightarrow
y^j_6$, valid for $i,j=1,\dots , 7$, guarantee that this polynomial will be of the form \be \beta
\sum_{i=1}^7 (y^i_6)^2.\ee Restoring $y^8_6 \neq 0$ gives a $(y^8_6)^2$ term and crossed terms
$y^i_6y^8_6$.
However, $y^8_6$ has been artificially isolated in the treatment of the GSO constraint: by isolating
other $y^i_6$'s and using the same arguments, we can show that our polynomial is of the from
\be \sum_{s\in{\bf 128}} (Q_a^{s}y^a_6)^2 = \beta \sum_{i=1}^8 (y^i_6)^2.\ee We obtain $\beta =
32$.

\noindent It is then straightforward to evaluate the sums encountered before. We obtain 
\be
M_T^2 = {2^3\over 4\pi R_6^2}\left(60\sum_{i=1}^{16} (y^i_6)^2\right),\ee and when coupling the Scherk-Schwarz
cycle to the helicity of the $E_8$ representation \be M_V^{(2)} ={2^3\over 4\pi R_6^2}
\left(-4\sum_{i=1}^{16} (y^i_6)^2\right).\ee\,

\noindent In the $N=2$ models, the orbifolding breaks $E_8 \rightarrow E_7 \times SU(2)$, under which the adjoint
representation decomposes as \be {\bf 248} \rightarrow ({\bf 133},{\bf 1})\oplus({\bf 56},{\bf
2})\oplus({\bf 1},{\bf 3}). \ee The Cartan generators of $E_8$ give the Cartan generators of $E_7 \times SU(2)$:
\be
\left(i\partial H_1,\dots,i\partial H_6,~i(\partial H_7 - \partial H_8)\right); ~~~i(\partial H_7+\partial
H_8).\ee Switching on arbitrary $y^1_6,\dots, y^7_6,y^8_6$, we compute the charges of the various
states step by step.

\noindent In the {\bf 133}, we have 7 neutral Cartan operators, 60 ladder operators in the $Adj(SO(12))$ subalgebra
\be e^{i(\pm H_j \pm H_k)},~~j\neq k \in \{1,\dots ,6\}, \ee 2 ladders in the $Adj(SU(2))$ \be e^{\pm
i(H_7-H_8)},\ee and 64 ladders in a spinorial representation \be e^{\frac i2(\pm H_1\pm\cdots \pm H_6
\pm(H_7-H_8))} \ee obeying a GSO condition. We see that $y^7_6$ has a particular role here.
The latter states have charges $\pm \frac 12$ under the first six Cartan generators, and charges $\pm 1$ under
the seventh. Using the same arguments as before, we see that the sum for the spinorial states is of the form \be
\alpha\left(\sum_{i=1}^6 (y^i_6)^2 + (2y^7_6)^2\right). \ee The polynomial we are looking for is
therefore \be \sum_{s \in{\bf 133}}(Q_a^{s}y^a_6)^2 = 20\sum_{i=1}^6 (y^i_6)^2 + 2(2y^7_6)^2+
16\left(\sum_{i=1}^6 (y^i_6)^2 + (2y^7_6)^2\right)= 36\sum_{i=1}^6 (y^i_6)^2 + 72(y^7_6)^2.\ee
Note that if we want to couple the Scherk-Schwarz cycle to the helicity of the $E_7$, we have
to compute also $\sum_s{\rm sign}(s)(Q_ay^a_6)^2$, where states in the spinorial representations of
the
$SO(12)$ subgroup contribute with a minus sign.
To get this sum we have to put a minus sign
in front of the {\bf 64} part, so that
$$ \sum_{s \in{\bf 133}}{\rm sign}(s)(Q_a^{s}y^a_6)^2 = 20\sum_{i=1}^6 (y^i_6)^2 +
2(2y^7_6)^2-16\left(\sum_{i=1}^6
(y^i_6)^2 + (2y^7_6)^2\right)$$
\be
\!\!\!\!\!\!\!\!\!\!\!\!\!\!\!=4 \sum_{i=1}^6 (y^i_6)^2 -56 (y^7_6)^2.\ee\,

\noindent For the ({\bf 56},{\bf 2}) representation,
we begin with the states with vertex operators \be e^{\pm H_i \pm H_7},~~~~e^{\pm H_i \pm H_8}.\ee The
corresponding $Q^s_ay_6^a$ are respectively \be \pm y^i_6 \pm (y^7_6 + y^8_6),~~~~\pm y^i_6 \pm (y^7_6 -
y^8_6).\ee
Therefore the sum for these states equals
$$
 \sum_s(Q_a^{s} y^a_6)^2 = 4 \sum_{i=1}^6 (y^i_6)^2 +24~(y^7_6+y^8_6)^2 +  4
\sum_{i=1}^6 (y^i_6)^2 +24~(y^7_6-y^8_6)^2
$$
\be
\!\!\!\!\!\!\!\!\!\!\!\!\!\!\!\!\!\!\!\!\!\!\!\!\!\!\!= 8 \sum_{i=1}^6 (y^i_6)^2
+48~((y^7_6)^2+(y^8_6)^2).\ee The remaining states to be considered have vertex operators \be
e^{\frac i2(\pm
H_1\pm\cdots \pm H_6 \pm(H_7+H_8))}. \ee For these, we get
\be 16\left(\sum_{i=1}^6 (y^i_6)^2 + (2y^8_6)^2\right).\ee Adding everything, we get the
final result for the representation:
\be  \sum_{s \in ({\bf 56},\bf 2)}(Q_a^{s} y^a_6)^2 = 24~\sum_{i=1}^6 (y^i_6)^2 + 48~(y^7_6)^2 +
112~(y^8_6)^2.\ee
If we couple to the $E_7$ helicity, we also have
\be  \sum_{s \in ({\bf 56},{\bf 2})}{\rm sign}(s)(Q_a^{s} y^a_6)^2 = -8~\sum_{i=1}^6 (y^i_6)^2 +
48~(y^7_6)^2 -16 ~(y^8_6)^2.\ee\,

\noindent For the {\bf 3} of $SU(2)$
the two states  \be e^{\pm i(H_7+H_8)} \ee have charges $\pm 2$. So we get
\be  \sum_{s \in ({\bf 1},\bf 3)}(Q_a^{s} y^a_6)^2 = 8~(y^8_6)^2.\ee\,

\noindent In the twisted sector, we encounter the representation ({\bf 56},{\bf 1}), whose sum is obtained
by switching off $y^8_6$ in the result obtained for the ({\bf 56},{\bf 2}) representation and by dividing the result by $2$. One then obtains,
\be  \sum_{s \in ({\bf 56},{\bf 1})}(Q_a^{s} y^a_6)^2 = 12~\sum_{i=1}^6 (y^i_6)^2 + 24~(y^7_6)^2,\ee
\be  \sum_{s \in ({\bf 56},{\bf 1})}{\rm sign}(s)(Q_a^{s} y^a_6)^2 = -4~\sum_{i=1}^6 (y^i_6)^2 +
24~(y^7_6)^2.\ee
We also encounter the ({\bf 1},{\bf 2}) representation, where the sum equals $2(y^8_6)^2$.

\noindent{\large \it Application to Models 3 and 4}\\
For model 4, we set $Q_R=Q_F + Q_H$. If we consider Wilson lines corresponding to the 16 Cartan generators
of $E_8 \times E_7\times SU(2)$, the result is
$$M_{T,V}^{2,(2)}  =  {4\over 4\pi R_6^2}\left[60~\sum_{i=9}^{16} (y^i_6)^2 + 36\sum_{i=1}^6 (y^i_6)^2 +
72(y^7_6)^2+24~\sum_{i=1}^6 (y^i_6)^2 + 48~(y^7_6)^2 + 112~(y^8_6)^2 + 8(y^8_6)^2\right]$$
$$ \!\!\!\!\!\!\!\!\!\!\!\!\!\!\!\!\!\!\!\!\!\!\!\!\!\!\!\!\!\!\!\!\!\!\!\!\!\!\!\!\!\!\!\!\!\!\!\!\!\!\!\!\!\!\!\!\!\!\!\pm {1\over 4\pi R_6^2}\left[32\left(12~\sum_{i=1}^6 (y^i_6)^2 + 24~(y^7_6)^2\right) +
128\left(2(y^8_6)^2\right)\right].$$
So we get
$$
\!\!\!\!\!\!\!\!\!\!\!\!\!\!\!\!\!\!\!\!M_T^2 = {1\over 4\pi R_6^2}\left( 240~\sum_{i=9}^{16} (y^i_6)^2 + 624~\sum_{i=1}^6 (y^i_6)^2+1248~(y^7_6)^2+736~(y^8_6)^2\right)
$$
\be M_V^{(2)} = {1\over 4\pi R_6^2}\left( 240~\sum_{i=9}^{16} (y^i_6)^2 -144~\sum_{i=1}^6
(y^i_6)^2-288~(y^7_6)^2+224~(y^8_6)^2\right) .\ee\,

\noindent For model 3, we set $Q_R=Q_a+Q_H+Q_{E_7}$. We get the same expression for $M_T^2$, while
$$M_{V}^{(2)}  =  {4\over 4\pi R_6^2}\left[4 \sum_{i=1}^6 (y^i_6)^2 -56 (y^7_6)^2 -8~\sum_{i=1}^6 (y^i_6)^2 +
48~(y^7_6)^2 -16 ~(y^8_6)^2 + 8(y^8_6)^2+ 60~\sum_{i=9}^{16} (y^i_6)^2\right]$$
$$ \!\!\!\!\!\!\!\!\!\!\!\!\!\!\!\!\!\!- {1\over 4\pi R_6^2}\left[32\left(-4~\sum_{i=1}^6 (y^i_6)^2 + 24~(y^7_6)^2\right) +
128\left(2(y^8_6)^2\right)\right]$$
\be \!\!\!\!\!\!\!\!\!\!\!\!\!\!\!\!\!\!\!\!\!\!\!\!\!\!\!\!\!\!\!\!\!\!\!\!\!\!\! = {1\over 4\pi R_6^2}\left( 112~\sum_{i=1}^6 (y^i_6)^2-800~(y^7_6)^2-288~(y^8_6)^2 + 240~\sum_{i=9}^{16} (y^i_6)^2\right) . \ee


\end{document}